\PassOptionsToPackage{table}{xcolor}
\documentclass[sigconf,nonacm]{acmart}

\usepackage{makecell}
\usepackage{rotating}
\usepackage{colortbl}
\usepackage{booktabs}  
\usepackage{changepage}
\usepackage{float}
\usepackage{natbib} 
\usepackage{subcaption} 

\usepackage{fancyhdr}
\definecolor{lightgray}{gray}{0.95}

\AtBeginDocument{%
  }

\setcopyright{rightsretained}
\copyrightyear{2025}
\acmYear{2025}
\acmDOI{XXXXXXX.XXXXXXX}
\acmISBN{978-1-4503-XXXX-X/18/06}

\newcommand{\killpunct}[1]{}

\makeatletter
\let\oldmaketitle\maketitle
\renewcommand{\maketitle}{%
  \oldmaketitle%
  \thispagestyle{plain}%
  \pagestyle{plain}%
}
\makeatother
\begin{document}

\title{Phraselette: A Poet’s Procedural Palette}

\author{Alex Calderwood}
\affiliation{%
  \institution{University of California, Santa Cruz}
  \city{Santa Cruz}
  \state{California}
  \country{USA}}
\email{alexcwd@ucsc.edu}

\author{John Joon Young Chung}
\affiliation{%
  \institution{Midjourney}
  \city{San Francisco}
  \state{California}
  \country{USA}}
\email{jchung@midjourney.com}

\author{Yuqian Sun}
\affiliation{%
  \institution{Midjourney}
  \city{London}
  \country{United Kingdom}}
\email{ysun@midjourney.com}

\author{Melissa Roemmele}
\affiliation{%
  \institution{Midjourney}
  \city{San Francisco}
  \state{California}
  \country{USA}}
\email{mroemmele@midjourney.com}

\author{Max Kreminski}
\affiliation{%
  \institution{Midjourney}
  \city{Berkeley}
  \state{California}
  \country{USA}}
\email{mkreminski@midjourney.com}
\orcid{0009-0002-6268-4033}

\begin{abstract}
According to the recently introduced theory of \emph{artistic support tools}, creativity support tools exert normative influences over artistic production, instantiating a \emph{normative ground} that shapes both the process and product of artistic expression. We argue that the normative ground of most existing automated writing tools is misaligned with \emph{writerly values} and identify a potential alternative frame---\textit{material writing support}---for experimental poetry tools that flexibly support the finding, processing, transforming, and shaping of text(s). Based on this frame, we introduce \textit{Phraselette}, an artistic material writing support interface that helps experimental poets search for words and phrases. To provide material writing support, \textit{Phraselette} is designed to counter the dominant mode of automated writing tools, while offering language model affordances in line with writerly values. We further report on an extended expert evaluation involving 10 published poets that indicates support for both our framing of material writing support and for \textit{Phraselette} itself.
\end{abstract}

\ccsdesc[500]{Applied computing~Arts and humanities}
\ccsdesc[300]{Human-centered computing~Empirical studies in HCI}
\ccsdesc[300]{Human-centered computing~HCI theory, concepts and models}
\ccsdesc[100]{Computing methodologies~Natural language generation}

\keywords{Creative Writing, Language Models, Poetry, Artistic Support Tool, CST, Material, Search}

\received{20 February 2007}
\received[revised]{12 March 2009}
\received[accepted]{5 June 2009}

\maketitle

\section{Introduction}

\begin{figure*}
    \centering
    \includegraphics[width=1\linewidth]{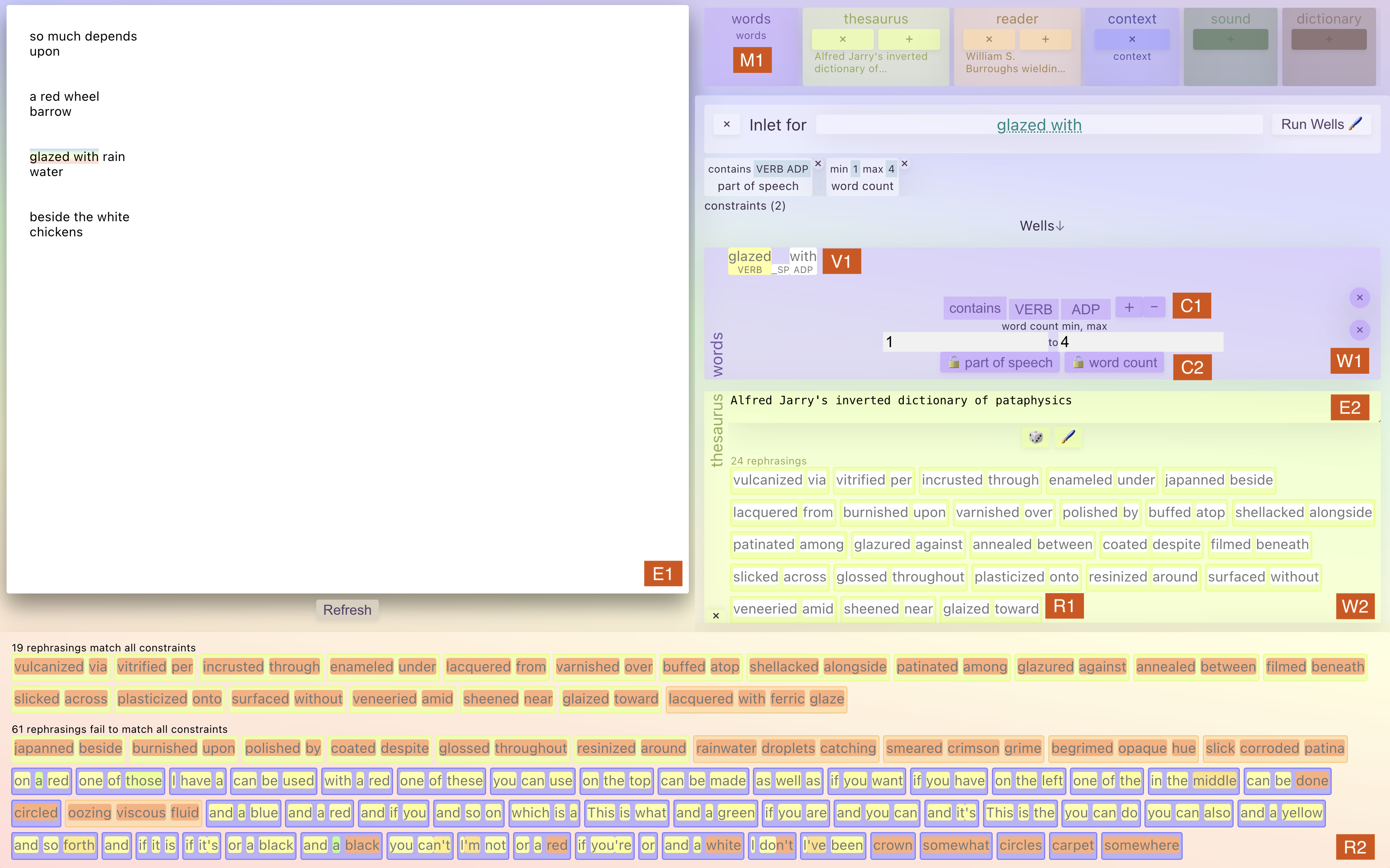}
    \caption{The \textit{Phraselette} interface. The editor (E1) allows writers to highlight text in order to create \textit{an inlet}, a site for text revision. Users activate \emph{phrasewells} with the side-scrolling menu (M1). Here the active phrasewells include a thesaurus (W1) which can be edited into different styles and roles with (E2) and a permanent well (W2) used for high level word constraints. Wells each provide some combination of \emph{rephrasings} (R1, R2), \emph{constraints} (C1, C2), \emph{insights} (Figure \ref{fig:reader}) and \emph{views} (V1)). A worked example is given in Section \ref{user-story} and a description of the affordances of wells in Table \ref{tab:wells_affordances}.}
    \Description{In the top left there's a text editor with certain spans of text highlighted for machine variation; in the top right there's a palette of many different phrasewells parametrized by input texts and other user-adjustable controls; at the bottom there's a list of many generated phrases, some of which match and some of which don't fully match the user's constraints.}
    \label{fig:interface}
\end{figure*}

As design researchers have introduced a variety of interfaces to support writing~\cite{lee_design_2024}, experimental creative writers have expressed misalignment between dominant writing support paradigms and their needs~\cite{booten_poetry_2021,calderwood_how_2018,UnmetNeeds}. A dominant paradigm in the design of computational writing support systems is the offloading of laborious tasks, with automation sometimes framed as reducing cognitive overhead in one domain to free up creative cognition in another. However, moving such processes from the writer's purview to the computer's can be damaging to the practice of creative writers for its tendency to dislocate texts from the process the text is undergoing. 

In poetry, and other creative writing, the `wordwork'---the trace of language from its origination (inspiration, provenance, discovery) through its revision (modification, appropriation, editing, alteration) into its realization (sharing, production, sociality \cite{morris_time_2006})---is of vital interest. It is understood that altering or automating one process affects others in rippling, nonlinear, and non-unidirectional ways \cite{hermansson_nomadic_2017,zhou_maintenance_2022}. The often prolonged engagement with the nuance of language along the chain of provenance is where discovery and learning take place \cite{sullivan_work_2013} and a work takes on the attributes that give it artistic meaning.

As we will explore, to design systems that facilitate artistic expression is to exert influence over the resulting art (Section \ref{artistic_support_tools}). Building tools for creative writing has elsewhere been likened to (musical) instrument design \cite{mitchell_toward_2021, wardrip_zot}, with every instrument lending its output a characteristic ``voice''~\cite{tanaka2006interaction,mitchell_toward_2021}. Meanwhile, in the context of poetry, some have begun to explore how such systems can be built in accordance with \textit{writerly} thinking or values (Section \ref{inkwell_writerly_design}).  We attempt to integrate this and other research on poetry interfaces \cite{booten_poetry_2021,gabriel_inkwell_2015} with a guiding theory of \textit{material} writing practice---involving textual recontextualization, modification, transmission, and representation. We refer to interfaces that support such practice as \textit{material writing support}, explained in Section \ref{material_writing_support}. These systems may explicitly provide textual material, and be designed with with the shifting interpretative landscape of the writer in mind.

\subsection{A contextual, constrained phrase thesaurus}

This paper outlines the design ethos for \textit{Phraselette}, a tool meant to---rather than offload creative cognition onto machine assistance---\textit{expand} the writer's possibility space by providing textual material only at sites where the writer requests it. It aims to introduce material to writers according to their specific intentions, integrating into existing creative writing practices by operating analogously to a traditional thesaurus. Like a thesaurus, it supports word search, and (unlike a thesaurus) does so with specificity to text context, with many avenues for creative control. Also unlike a traditional thesaurus, which supports only word-sized concepts, \textit{Phraselette} also allows search for multi-word phrases. 


Despite being framed in contrast to the automaticity of many existing large language model (LLM)-based applications, \textit{Phraselette} is aimed to expose the affordances of statistical language models and the decoding algorithms that accompany them. In various unusual ways, we attempt to provide users an ability to steer and look inside these models. The system provides many suggestions, offers cues about the model's state, emphasizes the poet's mental space over the text itself, enables a plurality of customizable search states, and aligns search criteria with a writer's notion of constraint. We put forward that certain affordances of language model technology, such as word retrieval capabilities and plain-text task specification, are, rather than analogous to or substitutive for writerly processes, able to provide a rich material work surface. These affordances provide opportunities for textual engagement and expansion parallel to both 1. the revision opportunities of traditional thesauruses (matching words, fulfilling constraints) and 2. writerly engagement with other texts (adaptation, pastiche).

\subsection{Contributions}

Our main contributions are as follows:

\begin{enumerate}
    \item We present design decisions demonstrating how language model-based search processes can align with writerly values and the existing processes of experimental poets.
    \item We introduce \emph{material writing support}, a frame to contextualize this design ethos within poetry and artmaking theory.
    \item We detail the design and implementation of \textit{Phraselette}, a system built under this design ethos.
    \item We report on an extended expert evaluation of the system (n=10 published poets), which suggests that \textit{Phraselette} had real-world utility without negatively impacting perceived ownership of written work.
\end{enumerate}


\section{Background}

\subsection{Artistic Support Tools} \label{artistic_support_tools}

\textit{Phraselette} is designed as an \textit{artistic support tool}. This follows the framing of \citet{li_beyond_2023}, who point out that creativity support tools exert ``\textit{power over} their users'' to construct a ``{normative ground}'' that structures users' ``ideas, goals, and intentions''. \citeauthor{li_beyond_2023} recognize that computational scaffolds ``further normative assumptions of what `good' art is'' through their production of artifacts and argue that for creativity support tools to succeed as artistic support tools, they should allow artists ``to define and redefine abstractions, flexibly navigate and compose parts of software tools, or resist and refuse tools altogether''. Our work asks `How might text generation affordances be built into an artistic support tool?' In Section \ref{design_box}, we outline how text generation can follow this theory's notions of \textit{vertical} and \textit{horizontal} movement in response to our position on materiality-based writing practice (Section \ref{material_writing_support}). 

One design paradigm we focus on is \textit{automaticity} and the mental model that writers adopt for a given writing tool. We aim to align a text generation system with writers' creative activity by exploding the unitary text generation process into an assortment of windows into various smaller subprocesses.
To this end, we limit generation length (search depth) and increase the number of considered alternatives (search width).
By allowing generative processes to produce only a single phrase at a time, we equate the unit of search with the scope of choices that we (as writers) make in our own creative writing practices. This `wide' search result paradigm, in which multiple generation processes work in parallel to offer up textual material, results in a wealth of opportunity for textual (re)interpretation and subsequent transformations.

\subsection{Writerly Values and Writerly Thinking} \label{inkwell_writerly_design}

\subsubsection{Inkwell} The poetry writing system \textit{Inkwell} \cite{gabriel_inkwell_2015} is a theoretical precursor to this work. \textit{Inkwell} allows users to craft Lisp-like poem templates. Specifically, the user writes some text directly and then, with user-specified constraints, \textit{Inkwell} generates other spans via search algorithms. For instance, the user might write ``the night'' followed by a query set: fill in a verb that connotes `dark' and rhymes with a noun that refers to a place. The system searches over the constraint space to produce `revisions' that satisfy these queries. \textit{Phraselette} was designed to support a similar concrete notion of constraint; like \textit{Inkwell}, it aims to supply a shared representation space that is familiar to poets and tractable to computational search.

\subsubsection{Writerly Thinking} Per Gabriel et al. \cite{gabriel_inkwell_2015}, \textit{writerly thinking} involves the "\textit{music} of words", subtexts, moods, connotations, metaphor, enjambment, surprise, subtle semantic play, and other ``craft elements'' relating to the form of the text. \textit{Inkwell} supports these considerations mostly by reasoning about concrete features of words, including synonyms, word senses, and emotional categorization. These, when composed in relation to each other by a search process and a poet's discernment, can aid that writer's process and produce novel \textit{association} or \textit{disassociation}. The success and novelty of \textit{Inkwell} is in its modeling of word-level abstractions that more comprehensively resemble a writer's concerns than other search systems. The ability to conjure a set of interrelated words that match writer-specified properties like connotation and tonality has a clear overlap with existing writer processes.

\subsubsection{Writerly Values} 

\citet{gabriel_inkwell_2015} back up their ability to represent writerly values by asserting that "one of the members of the InkWell team holds an MFA in Creative Writing". \textit{Phraselette}'s design is motivated in the same way---multiple coauthors have poetic practices that engage with digital materiality; \textit{Phraselette}'s design was conceived to provide search capabilities following these existing practices.

\subsubsection{Comparing Inkwell and Phraselette} Unlike \textit{Phraselette}, \textit{Inkwell} is a presented as co-creative system, wherein both human and computational processes are meant to exhibit creativity; to this point, \textit{Inkwell} has an autonomous mode, \textit{Phraselette} does not. Also unlike \textit{Phraselette}, \textit{Inkwell} requires its user to become versed in a dialect of Lisp, whereas \textit{Phraselette} does not demand any programming competence. We ask what it would look like to embed similar constraint search capabilities into an editor designed to provide suggestions not as singular answers to a search query but as interpretive material, available to a writer's subsequent exploration. Finally \textit{Inkwell} is a purely symbolic system, whereas \textit{Phraselette} is mostly based on statistical language model affordances (though also has symbolic search systems in various stages of development).

\subsection{Constrained, Computational, Lost} \label{lost}

In traditional metrical poetry, formal structure operates to scaffold a written work, in a limited sense to `generate' it \cite{fussell_poetic_1979}. Fixing a form, such a rhyme scheme or an aversion to the letter 'e', gives poets springboards and boundaries through which formal discoveries are made, meaning can arise, and play can unfold. In poetic theory, constraint flips the word's connotation of \textit{limit} to that of \textit{possibility} \cite{fussell_poetic_1979,thomas_oulipo_1988}. The pursuit of novel forms of constraint is an ancient tradition, and common in experimental formal literary practices such as that of the Oulipo \cite{thomas_oulipo_1988}.

In computer writing \cite{kirschenbaum_track_2016,kirschenbaum_materiality_2001}, the active representation of a constraint might have the purpose of limiting a computational search space, with the expectation that the discovery of new linguistic patterns, phrases, and words might assist writers fulfill a presumed intent (Gabriel et al.'s \textit{associative writing}).

However, as Kreminski observes in an essay on \textit{lost poetry} \cite{kreminski_computational_2024}, computer and human interpretations of textual meaning cannot be truly shared. 
Borrowing Parrish's framing of computational search processes as ``semantic space probes'' \cite{parish_exploring_nodate} that can discover linguistic patterns humans wouldn't otherwise find, Kreminski argues that disjunctions between what a computational system `sees' and the interpretations that human writers form from machine process are a verdant ground where new ideas emerge. Conducting a computational search is to wed a set of necessarily disunited or alien representations: the human and computer views. Search might then lead to language that has "unexpected agreements" with a poet's writing, perhaps leading them to ``new ways of representing language'' \cite{kreminski_computational_2024} (related to Gabriel et al.'s \textit{dissociative writing}).

\subsection{Material Writing Support} \label{material_writing_support}

\subsubsection{What is Material Writing?} In simple terms, we define material writing as writing that engages with its own representation or the processes through which it arises \cite{morris_time_2006, cayley_time_2006,kirschenbaum_materiality_2001}.
In this section, we build up our claim that text-generation technologies, including those designed for constraint, can support material writing practice. 

Variations of this broad view of writing are now commonly taught, and follow the dream practices of Yeats \cite{holdeman_introduction_nodate} through to avant-garde writers: the appropriative writing of Kathy Acker \cite{doherty_kathy_2022} and more contemporary writers such as Tom Comitta whose conceptual text ``The Nature Book'' pulls nature passages from dozens of texts, and David Shields'  interwoven hypernarrative, ``How Literature Saved My Life'' in a style of collage or montage. These writers' relationship to prior texts creates the conditions for new work. We use N. Katherine Hayles' essay \textit{The Time of Digital Poetry: From Object to Event} as a guiding theory text that establishes the relationship between experimental poetry and the notion of materiality \cite{morris_time_2006}. 






\subsubsection{Text in motion} Hayles quotes the poet and critic Loss Pequeño Glazier, who writes ``materiality is key to understanding innovative practice'' \cite{morris_time_2006}. Within this new media theory, a digital text's materiality can be first understood as the substrate that carries language to its audience; Hayles calls this its "production/performance", referring to the page of a book, the form of an email's presentation, or the choreography of a moving poem. Materiality invokes the notion that poets engage with a system of interfaces, persons, and other texts, to give rise to a text. Digital texts specifically exist as \textit{processes} rather than fixed things, according to Hayles. Digital innovations in poetic practice, such as moving type poems, have made obvious the preexisting truth that language is in motion, existing as a series of transcriptions, translations, and transcodings \cite{morris_time_2006}. 

This view implies that to support innovative poetry and literary writing practice, we must attend to the manipulations that text undergoes, the involvement of ``proprioceptive projection'' (ibid.), representations, and processes which we have earlier referred to as the text's `chain of provenance'. Hayles quotes Glazier: ``the defining characteristic of literary language is the impulse to investigate its conditions of possibility''. These conditions are the text's materiality. Matthew Kirshenbaum also observes that attention to digital materiality leads us to a view of production and process. Quoting McGann, Kirshenbaum writes, ``The relevance of that `complex network of people, materials, and events' that lies behind textual production is only amplified in electronic settings'' \cite{kirschenbaum_materiality_2001}.

\subsubsection{Interpretation} The final extension we need to make to get to material writing support is the connection between materiality and interpretation (via the relationship of materiality and process).

We observe that each processual link in poetic production affords an opportunity for a poet to re-interpret a text. Consequently, each transformation, borrowing, and recontextualization signifies a potential change in the poet's design object. It may be best to consider the design object that poets engage with to be not the text but the \emph{interpretation} of the text. Each subtle transformation and manipulation, an added word or a blurred relationship between words, has the chance to take on an important role for the poet's conceptual process. 

Textual provenance thus operates within the writer's mental space, and our view of materiality thus concerns itself with understanding the importance of interpretation at sites of textual transformation. The artist Giselle Beiguelman's conception of \textit{nomadic poetry} echoes this focus: for Beiguelman, text clippings and ad billboards have the power to transform interpretation and mental space; ``second-generation originals and media info-bodies'' become sources of poetic engagement \cite{nomadic_poetry}.

\subsubsection{Support} \label{support-defn} We define \textit{material writing support} as that concerned with the interpretive spaces that texts move through as manipulations, additions, and processes unfold. As nomadic poetry emphasizes engagement with text as it moves from the world to the mind, \textit{material writing support} should facilitate the play of text across sources and interpretations. It should seek to support the design of interpretation through the cross-play of text.

The design recommendations of \textit{material writing support} tools aim to bring attention to the relation of a digital system and the writer's interpretive landscape. 
Kreminski's framing of lost poetry (Section \ref{lost}) suggests that AI systems may be able to facilitate innovative experimentation through the disjunctions or surprising interpretive conjunctions that arise when text moves from an AI system to the mind.
Meanwhile, \citet{buschek_collage_2024} sketches out a collage-inspired view of how user interfaces might incorporate material from multiple sources via intentional \textit{fragmentation} (or \textit{de-fragmentation}) of text views.
Extending these perspectives, \textit{material writing support} might:


\begin{enumerate}
    \item Produce textual material and desired perturbations in landscape (\textit{Phraselette}'s focus)
    \item Highlight and contextualize relationships between textual elements 
    \item Explore the operation of text, provide interpretations for revision
    \item Permute existing works
    \item Widen or reframe texts, move them into new conceptual spaces
\end{enumerate}

\subsubsection{Interaction Metaphor and Mental Model} \label{background_mental_model}

The conditions that lead writing tools to successful integrate within real world art-making practice is somewhat understudied in the HCI literature surrounding writing tools. \citeauthor{lee_design_2024}~\cite{lee_design_2024} describe how the \textit{interaction metaphor} of a writing assistant shapes design considerations and frame the user's relation to the system. They identify dimensions of \textit{Agency}, \textit{Ownership}, \textit{Integrity}, \textit{Trust}, \textit{Availability}, and \textit{Transparency}, with each ``influencing the long term perceptions of the system'', and thereby its utility and adoption.

 
 A system that is viewed as altering perceptions of a work's derivation without making that alteration legible (affecting the Ownership and Agency dimensions \cite{lee_design_2024}) might be detrimental to a confessional poet interested in expression and directness, while a system that misrepresents the views of a marginalized group will be engaged with differently by artists interested in representation and power. The mental model (used interchangeably with `interaction metaphor' in \cite{long_not_2024}) occupied by material writing support tools will always be relational to the other material circumstances of a work.

\section{The Normative Ground of Text Generation} \label{design_box}


The normative ground that autonomous writing systems exert is still beginning to be understood. It can be cast in terms of the predilections of language model outputs, mostly attributed to the training regimes and data sources used to produce a system's weights, as well as the algorithmic processes used to produce texts. We focus on how we can re-examine these algorithms and their interface dynamics in order to move the normative ground exerted by the interface.

We synthesize the normative ground of many autonomous writing systems as:

\begin{itemize}
    \item \textit{Wholesale}: Generates longform text to accept or reject as one unit.
    \item \textit{Singular}: One (or few) suggestions are provided.
    \item \textit{Inscrutable}: The system's operations are obscured. Aside from a few parameters (temperature, etc.) .
    \item \textit{Authoritative}: The interface places text directly in the document.
\end{itemize}


It might be argued that autonomous writing systems like ChatGPT can be used in the material support mode described in Section \ref{support-defn}. But we claim that these interfaces, partly for the mental model they induce (Section \ref{background_mental_model}), produce a normative ground (Section \ref{artistic_support_tools}) not conducive to the interpretive moves writers are looking to make in the production of their work \cite{booten_poetry_2021,li_beyond_2023}. While autonomous writing \cite{robinson_speculative_2022} does produce material, it is somewhat random, singular, and decisive, rather than purposeful, open, and plural.

As \citet{li_beyond_2023} detail, artistic support tool designers can enable \textit{horizontal movement} by creating interoperability between systems, allowing users to move their work through different toolings and workflows, and giving the creator decisive authority. Our framing of materiality is in line with the notion of horizontal movement in the way that it is designed to facilitate varied interpretive affordances to the poet. That is, the system provides \textit{horizontal movement} in occupying a point in design space characterized as:

\begin{itemize}
    \item \textit{Piecemeal}: Providing small chunks of text for the writer to integrate into a cohesive whole.
    \item \textit{Multiple}: Providing many suggestions.
    \item \textit{Transparent}: Providing insights and interpretable cues that indicate the state of the model.
    \item \textit{Suggestive}: Placing emphasis on the poet's mental space rather than the text itself, allowing their interpretations to decide how the text should be manipulated.
\end{itemize}

\citeauthor{li_beyond_2023} say tool designers can enable \textit{vertical movement} by creating flexible systems that allow users to move through abstractions. They indicate that designers should be wary of rigid abstractions, which our constraints might be accused of being. We justify it as necessary to provide our \textit{horizontal movement} and argue that creating vertical movement sometimes entails moving abstractions towards a space legible to users. Our system provides \textit{Vertical Movement} by:

\begin{itemize}
    \item Providing custom handles like free text entry as a way to influence text search/generation
    \item Providing search constraints legible to writers
\end{itemize}

\section{System Description}

\subsection{System Design}

\begin{table*}[h]
\centering
\begin{tabular}{|p{3cm}|p{10cm}|}
\hline
\textbf{Affordance} & \textbf{Description} \\ \hline
\textit{Rephrasings} & These are rewrite suggestions made by the Wells. Rephrasings appear within the Well UI element and are also aggregated to a window on the bottom of the screen, colored according to the Well in which they originated. \\ \hline
\textit{Constraints} & Some Wells allow the writer to add constraints that reduce the search space of its own generation capabilities as well as the generation of other Wells. \\ \hline
\textit{Insights} & Additional information that a Well provides. In some cases, insights are textual: such as feedback generated according to a custom role. Insights can also be graphical; such as a word likelihood histogram. Insights function partly as mechanisms for the user to ``see what the machine sees'', and form a closer relationship to the generative system responsible for rephrasings. \\ \hline
\textit{Views} & Some Wells provide a unique \textit{view} on the selected text. The view may be thought of as a lens through which the document may be annotated. These Wells attach additional data to the words (tokens) for all rephrasings in the document. Wells that provide a view display it for the current selection, (shown in Figure \ref{fig:words}), and to the expanded display shown when rephrasings are hovered (Figure \ref{fig:rephrasing}). \\ \hline
\end{tabular}
\caption{Affordances provided by Wells in the Phraselette system.}
\label{tab:wells_affordances}
\end{table*}

\textit{Phraselette} is modular, supporting a variety of text generation strategies and modes of textual constraint. Disparate text generation/search algorithms, models, and forms of data representation (Table \ref{tab:data-types}) can thus be used in concert, each with different strengths and use cases that writers can tailor for their own purposes.

The system connects text generation strategies, constraints, and insights into the conceptual unit of the \emph{phrasewell} (Table \ref{tab:wells_affordances}). This early design decision was made based on the observation that introducing a new model or text generation strategy often also introduces a accompanying analytical lens: the supporting data used by that generation strategy to arrive at suggestions.

Likewise, new data sources and analytical lenses usually bring with them new generative capabilities. Associating generation strategies with mechanistic explanations of textual provenance unites our goal of making the system's generative capabilities interpretable with our goal of giving users novel insights into a text's production.



\subsection{Extensibility} 
\label{extensible}

Linking these affordances into phrasewells also improves modularity and extensibility. For instance, a programmer might like to introduce a WordNet-style conceptual ontology to \textit{Phraselette}'s text representations. Wells are designed with object oriented principles, so the extension can be made by implementing a new phrasewell class. Overriding the class means defining the generation and text representation affordances detailed in Table \ref{tab:wells_affordances}.

\begin{table}[h]
\centering
\begin{tabular}{l|l|c}
\textbf{Data Type} & \textbf{Constraint} & \textbf{Implemented} \\
\hline
Symbolic & phonemes (pronunciation) & $\checkmark$ \\
Symbolic & part of speech & $\checkmark$ \\
Continuous & word likelihood & $\checkmark$ \\
Discrete  & word count & $\checkmark$ \\
Discrete & syllable count & $\checkmark$ \\
Vector & word embeddings & $\circ$ \\
Key-value pairs & ontological relations & $\circ$ \\
\end{tabular}
\caption{\textit{Phraselette} supports many types of text constraints. Each supported example pairs a UI widget for specifying a desired textual attribute (e.g., a range of acceptable syllable counts) with computational method(s) for steering generators toward that goal.}
\label{tab:data-types}
\end{table}

\begin{figure}
    \centering
    \includegraphics[width=0.75\linewidth]{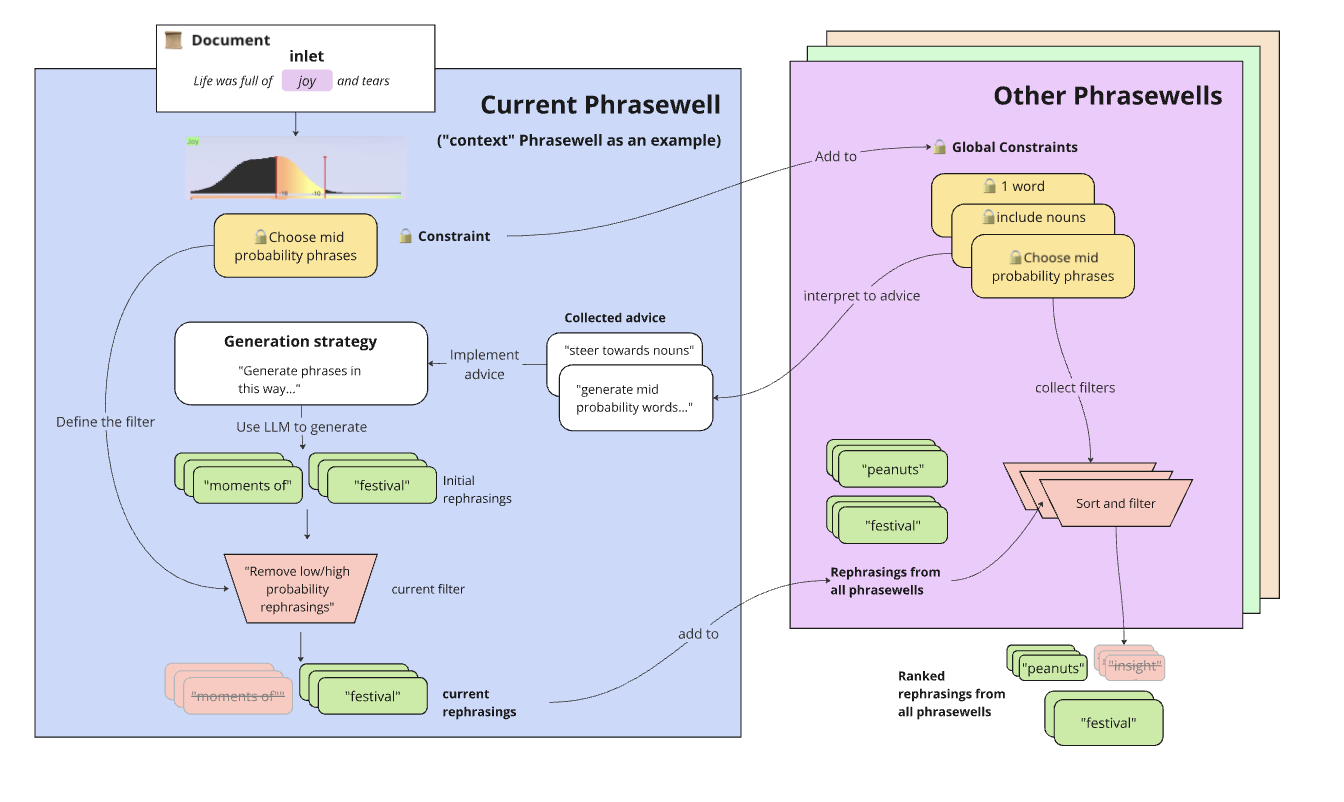}
    \caption{Flow diagram demonstrating the operation of a single phrasewell. Rephrasings are produced and internally scored using active constraints, before being mixed with the rephrasings from other phrasewells.}
    \Description{Text flows from a highlighted span, or "inlet", in the input document down into each active phrasewell; the phrasewell then uses its internal generation strategy, potentially parametrized by user-supplied inputs and advice from other active phrasewells, to generate some number of viable rephrasings of the text in the inlet. Every phrasewell then contributes constraints to the filtering and sorting of the pool of potential rephrasings generated by the set of all active phrasewells.}
    \label{fig:architecture}
\end{figure}

\subsection{Working with Phraselette} \label{user-story}

\subsubsection{A User Scenario} To illustrate how \textit{Phraselette} works in practice, we will demonstrate its use for revision. We present a user scenario: A poet, Charlie, wants to continue writing a series of poems that critiques the naivete of modernism. Charlie often works by modifying other texts, and here begins with William Carlos Williams' ``The Red Wheelbarrow''. Charlie opens the editor and pastes the poem (Figure \ref{fig:interface}).

Charlie recognizes the phrase ``glazed with'' is doing a lot to guide the temperament of the scene. Charlie highlights it in the editor and selects \textit{Create Inlet}. The system will now maintain this highlighted span (or \textit{inlet}) as they work, and aim to produce revisions of the selected text.

The interface displays available text transformation tools, the phrasewells (or simply ``wells''), in a top-righthand-side menu (Figure \ref{fig:interface}, \textbf{M1}). Each well represents a different approach to text transformation. Charlie activates a `reader' Well by clicking its (+) icon. It appears in the right hand column alongside an initial `words' Well\footnote{not pictured in Figure \ref{fig:interface}, clipped by the other wells}. 

Charlie wants this descriptive phrase to evoke a whimsical and actively alternative, technical tone but but they also know that it should describe the behavior of water.

Charlie cycles through the included reader archetypes by pressing the Well's die icon until it reads ``William S. Burroughs'' In addition to this reader, Charlie adds a thesaurus Well (\textbf{W1}), which they know to be less impacted by the text surrounding the highlighted inlet; they hope this will produce more contrast with the current language. To describe how this thesaurus should behave, Charlie writes ``Alfred Jarry's inverted dictionary of pataphysics'' to target an appropriately whimsical and parodic style (\textbf{E2}). Finally, they add a context Well\footnote{also clipped} to get a sense of what words would be most likely to appear in the place of ``glazed with'' (which they want to avoid).

\subsubsection{Phrasewells}

Phrasewells are the main way that users interact with \textit{Phraselette}. When triggered, wells can provide a handful of unique affordances, which we describe in Table \ref{tab:wells_affordances}.

Section \ref{well_types} gives the types of Wells supported during the user study. Each of these use language models, but we should note that this framework also allows wells to incorporate other text generation strategies; BabelNet~\cite{BabelNet} to jump between word associations, for instance.

\subsubsection{Running Wells}

Now that Charlie has assembled an assortment of Wells, each with slightly different foci, they can generate alternate rephrasings for the selection either by triggering individual Wells or by running all active Wells simultaneously. 

\begin{figure}
    \centering
    \includegraphics[width=0.3\linewidth]{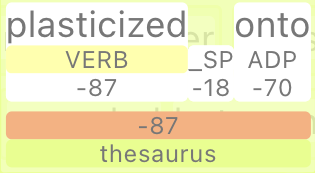}
    \caption{Hovering over a rephrasing expands the view data contributed by active Wells. Here, `Word' and `Context' Wells provide part of speech and (logarithmic) word probabilities.}
    \Description{A single rephrasing consisting of the text "plasticized onto", which is visibly divided into three tokens (the verb "plasticized", a space, and the preposition "onto"). Below each token, a part-of-speech tag and a logarithmic probability score are both visible. Below the whole phrase, an overall logarithmic probability score for the phrase and the name of the phrasewell that contributed this rephrasing ("thesaurus") can also be seen.}
    \label{fig:rephrasing}
\end{figure}

Charlie decides to move towards more active language in order to evoke a sense of critique. In a permanent Well that deals with word-level constraints, Charlie clicks the lockbox emoji, and using the constraint selection interface for part of speech (symbolic data per Table \ref{tab:data-types}), changes to a Verb-Adverb structure (\textbf{C2}). 

After Charlie runs the Wells, they each provide various forms of text or other insights (Table \ref{tab:wells_affordances}). Hovering over any rephrasing shows all the view data associated with it (Figure \ref{fig:rephrasing}). Commonly, the user will see the insights and ideas on the screen and simply make note of the new diction or come up with a different idea to prod towards. Clicking on a rephrasing will place it in the current inlet, but a common use case is to absorb the responses given by the Wells and manually make a change to the current inlet or elsewhere. For \textit{Phraselette}, success doesn't always entail direct user acceptance of machine suggestions---noting a related word or concept often leads to revisions across the text, and may lead to knock-on effects to structure and concepts \cite{zhou2024ai,roemmele2021inspiration}. 

Charlie sees that \textit{Phraselette} has elevated the rephrasings ``vulcanized via'' and ``vitrified per'' (Figure \ref{fig:interface} \textbf{R1}). They need to look up what vitrified means (perhaps consulting the dictionary Well for an approximate definition)---to convert into glass by heat. They hover over the rephrasing, noting that the context Well has satisfyingly provided a low probability for the word, and click on it, replacing ``glazed with'' in the text. Noting the play of ``vulcanized via'', they edit the stanza to: ``vitrified via rain / water''.

They note that vitrification via water entails that the water must be \textit{hot}, so they shift focus and highlight the (red) ``wheel barrow'', creating a new inlet for the phrase without removing the previous inlet. Clicking `Run Wells' restarts the search on this inlet with the same active Wells as before, allowing them to apply the same palette of word search tools to the new context.


\section{Types of Phrasewells} \label{well_types}

\subsection{Basic (`Words')}

\begin{figure}[H]
    \centering
    \includegraphics[width=0.8\linewidth]{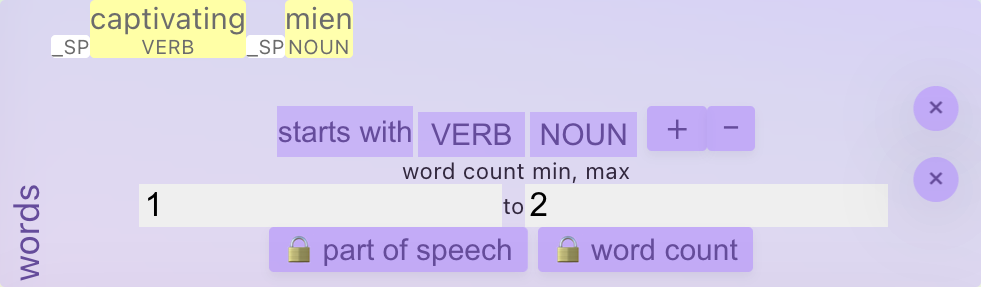}
    \caption{The most basic phrasewell is used to create simple word-based constraints}
    \Description{Visible controls for the words phrasewell include a selector for how to apply part-of-speech constraints (currently set to "starts with"); a sequence of target part-of-speech tags (currently set to "VERB NOUN"); and inputs for a word count minimum and maximum (currently set to 1 and 2 words respectively). Buttons for adding and removing constraints are also visible.}
    \label{fig:words}
\end{figure}

\subsubsection{Operation} The `Words' Well is the only phrasewell that cannot be disabled. It allows users to constrain text by word count and part of speech (POS) tags. As with other symbolic constraints, the user can specify sequences that rephrasings should contain, begin/end with, or have in order. This Well does not generate rephrasings or insights.

\subsubsection{Advice and Scoring} 

Constraints have two roles during rephrasing generation:

\begin{enumerate}
    \item To produce \textit{advice} to the search algorithm (information that guides and directs the search)
    \item To \textit{score} the rephrasings produced by the Well, for sorting
\end{enumerate}

Because each type of Well that produces rephrasings has a unique search algorithm, it is up to each Well to determine how to satisfy each constraint (Figure \ref{fig:architecture}). We refer to a Well modulated by a given constraint as receiving `advice' from the constraint.

For instance, the word count constraint provides advice to some wells by directly changing parameters to the beam-search process used in rephrasing. For others, it adds guidance to the prompt in plain English, e.g., ``aim to produce between 1 and 4 words''. The same is true for POS constraints, which provide advice both in the form of prompt injections and (experimentally) a beam search heuristic to guide the sampling process.


\subsection{Thesaurus}

\subsubsection{Operation}

\begin{figure}[H]
    \centering
    \includegraphics[width=\linewidth]{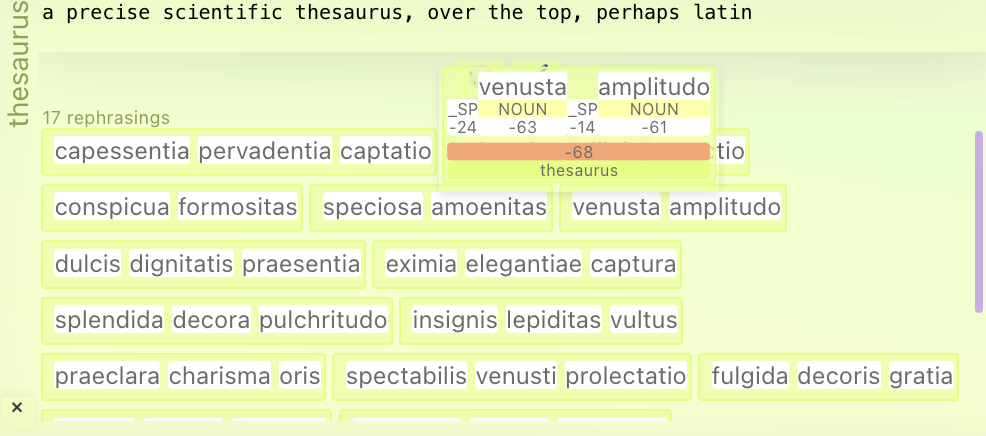}
    \caption{The thesaurus phrasewell allows custom plain-text guidance over desired word properties.}
    \Description{The only visible control for the thesaurus phrasewell is a text input box specifying the kind of thesaurus to simulate (currently set to "a precise scientific thesaurus, over the top, perhaps latin"). Below this input box, several rephrasings generated by the phrasewell are visible; one, "venusta amplitudo", is expanded, showing extra information about the rephrasing like that in Figure 3.}
    \label{fig:enter-label}
\end{figure}

The thesaurus Well allows users to enter plain-text descriptions for the thesaurus the user would like to generate the rephrasings.  Built-in thesaurus descriptions include:

\begin{itemize}
    \item ``the thesaurus James Joyce used for Ulysses''
    \item ``a thesaurus of homophones and near homophones (words having the same pronunciation but different meanings, origins, or spelling)''
    \item ``a romance novel's lexicon''
\end{itemize}

\subsubsection{Implementation}

The thesaurus phrasewell implements constraints primarily by algorithmically constructing LLM prompts that contain instructions to adhere the set of active constraints. 

The LLM prompt includes the Inlet's selection text, but does not include the context surrounding the text under revision. This is unlike the other Wells, which do take into account the surrounding context. Restricting the information used to inform generation will change the bent of the rephrasings. The decision showcases that a multiplicity of diverse generation strategies, each with different aims and affordances, can be used in concert.

\subsubsection{Background}

The thesaurus Well owes some design inspiration to \cite{gero_how_2019}, which trained a word2vec model~\cite{word2vec} on stylistic corpora to come up with similar words. \textit{Phraselette}'s thesaurus similarly uses a machine learning model to produce rephrasings; it therefore has a different `timbre' (to use the earlier instrument metaphor) but a wider range, because it infers style from user-editable prompts rather than pre-computing a fine-tuned word2vec model for each offered style. This Well excels at neologisms and providing unusual phrases.

\subsection{Reader}

\begin{figure}[H]
    \centering
    \includegraphics[width=1\linewidth]{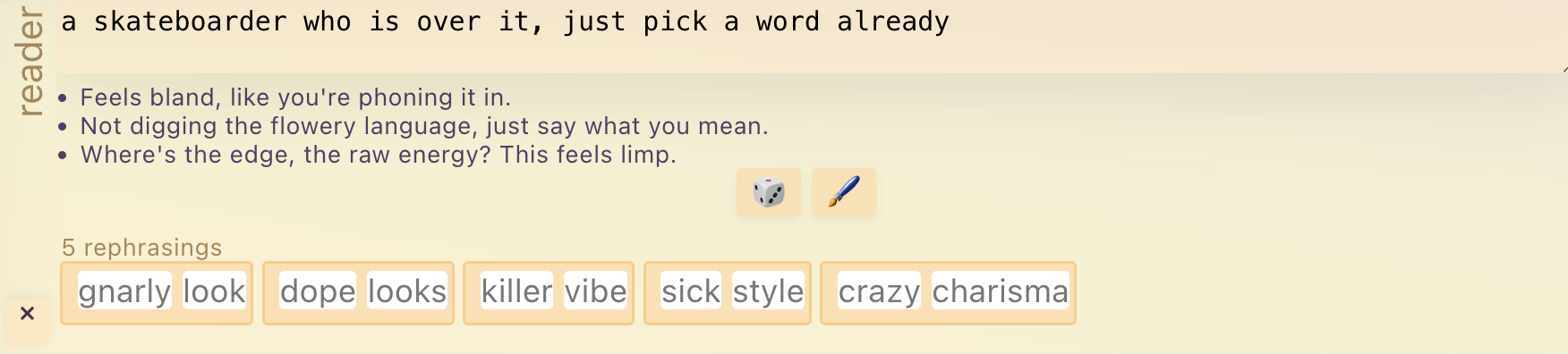}
    \caption{The reader phrasewell gives poets the ability to `cast' an imagined persona into a role of reader to provide feedback and revision material.}
    \Description{The only visible control for the reader phrasewell is a text input box specifying the persona of the reader to simulate (currently set to "a skateboarder who is over it, just pick a word already"). Below this input box, three bullet points of commentary generated by this persona on the currently highlighted text can be seen; the commentary includes sentences like "Not digging the flowery language, just say what you mean." Below the commentary, a handful of rephrasings generated by this phrasewell are also visible; these include "gnarly look", "killer vibe", "sick style", and so on.}
    \label{fig:reader}
\end{figure}

\subsubsection{Operation} Like the thesaurus, the reader phrasewell allows users to describe the attributes of a desired reader persona in plain text. The user can rotate through predefined persona descriptions, or introduce their own. Built-in reader descriptions include surrealist writers (``Tristan Tzara, the Dadaist poet'') and archetypes (``a literary critic...''). This description is then used to construct an LLM prompt which also includes the document's context and the current selection text.

Upon triggering the reader Well, it first provides up to three bullet points that are generated to characterize a critique or positional reading of the current selection, given the perspective of the reader archetype in the provided description. These insights (or `reader responses') are passed into a second prompt, which instructs an LLM to produce rephrasings in accordance to this advice. This two-step process is inspired by autodebugging \cite{kelly_there_2023} and the improvement of LLM writing via edit models~\cite{AIWritingSalvaged}. It typically provides 5-12 rephrasings. 

\subsubsection{Background}

The reader phrasewell responds to various calls for computational readings to provide alternate perspectives, represent marginalized voices, or help understand diverse viewpoints during the analysis and production of text \cite{wordcraft,moretti_graphs_2005,ramsay_reading_2011,kreminski2024dearth}.

Its name references \textit{The Readers Project} \cite{howe_reading_nodate}, a conceptual art project that visualizes how simple text processing algorithms traverse texts. This project's various readers operate both similarly and dissimilarly to human readers, illustrating how reading modes can alter the interpretive affordances of textual material as they generate alternative traces of their reading path.

There have been many calls to use LLMs to represent diverse human viewpoints, hoping to amplify marginalized voices or give writers stylistic cues into other writing styles \cite{coenen_wordcraft_2021,kreminski2024dearth}. 
Diverse LLM-generated interpretations may also bring about what Kreminski calls an ``\textit{abundance of the author}, in which every word of a piece has been considered more carefully and from more different angles than the author could otherwise manage or afford'' \cite{kreminski2024dearth}. Yet, LLM interfaces built with this goal often fail to find true parity in form, style, or content with the person(s) being represented~\cite{BiasRunsDeep,Persona-L}. \textit{Phraselette} hasn't solved this problem, but partially sidesteps it by offering many \textit{possible} rephrasings in the manner of a thesaurus; \textit{Phraselette}'s suggestions thus do not each need to be perfectly apt to provide value via inspiration and honing language. As our findings suggest, even a single compelling word or phrase among alternatives can prove the system's value to creative writers.


One study participant, P7, suggested that a reader phrasewell can be thought of as a ``persona''. This aligns well with Jeremy Douglass' classroom activities that informed our design, in which writing students hone language by prompting LLM's with ``persona and roles'' \cite{douglass_writing_2023, usc_future_of_writing_symposium_jeremy_2023}.

\subsection{Context}

\begin{figure}
    \centering
    \includegraphics[width=0.7\linewidth]{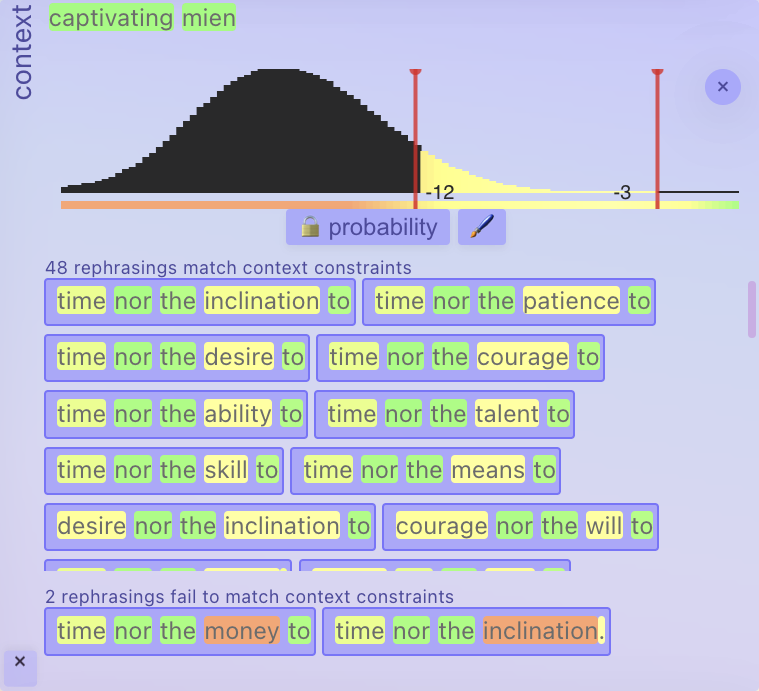} 
    \caption{The context phrasewell generates within the bounds of the probability distribution set by the user.}
    \Description{The context phrasewell is controlled by a visual curve representing a probability distribution; the user can adjust the endpoints of a sampling range on this curve to adjust from what parts of the probability distribution rephrasings are allowed to be drawn. Below this, many very similar rephrasings are also visible.}
    \label{fig:context-well}
\end{figure}

\subsubsection{Operation}
 This phrasewell surfaces the top 20-50 rephrasings, using a custom beam-search scoring algorithm \cite{katnoria_visualising_2020,freitag_beam_2017} which samples from among the most likely continuations to the text immediately preceding the Inlet, taking no account of the Inlet's highlighted text. 

The context phrasewell provides insights in the form of a histogram that shows the log-likelihood distribution of possible rephrasings according to the model. It also provides probability constraints, which can added by dragging range boundaries over the distribution. Doing so will attempt to limit the rephrasings (across all Wells) to those falling within this range.

\subsubsection{Background}

The context phrasewell might be the purest form of our commitment to rethinking the affordances of LLMs. Despite utilizing statistical techniques (a single beam search run on a past-generation language model), we can expose the `thesaurus-like' properties of LLM's to users who value a peek into the `normativity' implied by directly examining the weights of a foundation model. This Well also provides users with a means to explicitly \emph{exclude} the likeliest possible phrases from consideration, potentially mitigating the homogenization effects~\cite{PadmakumarHe,DoshiHauser,HomogenizationEffects,TowardWesternStyles,PredictableWriting} that LLM support for creative writing may introduce.

During implementation, we discovered \citet{roush_most_2022} seem to have achieved success with a project that follows a similar design and implementation of `hard constraints' used to inform LLM beam search results.

\subsection{Sound}

\begin{figure}[H]
    \centering
    \includegraphics[width=0.7\linewidth]{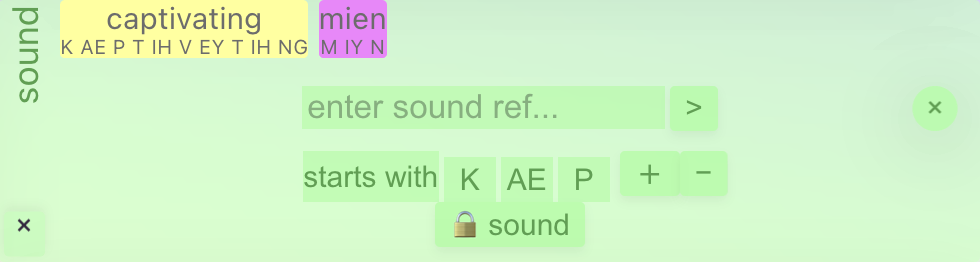}
    \caption{The sound phrasewell provides a well understood search space (word sounds) to poets via the sound constraint.}
    \Description{The sound phrasewell is controlled by a "sound ref", which can be inferred from an input span of text or manually entered as a sequence of raw phonemes (currently set to "K AE P"); there's also a mode selector for setting how this constraint will be applied (currently set to "starts with"). At the top of the phrasewell, the inlet text it's currently operating on is annotated with information about how it's pronounced: "captivating mien" is rendered as "K AE P T IH V EY T IH NG M IY N".}
    \label{fig:sound-well}
\end{figure}

\subsubsection{Operation}

The Sound Well is uniquely interested in providing constraints over the phoneme space (speech sounds) of words. Activating the Well introduces a phoneme-generation step to the rephrasing pipeline. Users can then constrain the sound space of rephrasings to match parts of a reference pronunciation, e.g., to start with a `k` sound or rhyme with the end of the current selection. 


\subsection{Dictionary}

\begin{figure}[H]
    \centering
    \includegraphics[width=0.7\linewidth]{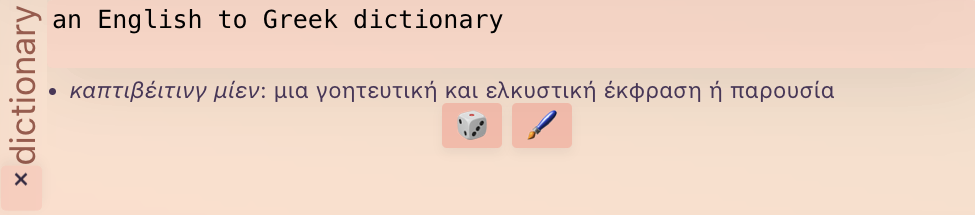}
    \caption{The dictionary phrasewell can be used for translations, as well as stylized responses (such as a nonsense dictionary). }
    \Description{The dictionary phrasewell, like the thesaurus and reader phrasewells, is controlled by an input text box describing what kind of dictionary to simulate; here it's set to "an English to Greek dictionary" for translation purposes. As expected, the generated definition text for the inlet text (below the dictionary well's input text box) is rendered in Greek.}
    \label{fig:dictionary-well}
\end{figure}

\subsubsection{Operation}

Similar to the thesaurus, this Well allows users to provide a text description to guide the system in producing machine generated definitions (\textit{insights}). The Well takes into account the context around the current Inlet as well as the selection text itself to produce the stylized or otherwise purposeful definitions.

\subsubsection{Background} 

This Well was included partly to demonstrate the breadth of possibilities provided by \textit{Phraselette} framework. We noticed that the thesaurus can often provide new coinages as well as unknown phrases and wanted to match them with possible definitions. 

\section{User Study}

\subsection{Research Questions}

We conducted an \textit{extended user study} (Section \ref{extended-user-study}) with 10 experts (recently published poets) over the course of roughly two weeks of use. We asked:

\textbf{RQ1}: Do writers find that the system's combination of text generation strategies and constraints useful in their writing practice?

\textbf{RQ2}: How does the mental model that writers form of the system affect their writing? (Section \ref{mental})

\begin{figure*}
    \centering
    \begin{minipage}{0.48\textwidth}
        \centering
        \includegraphics[width=\linewidth]{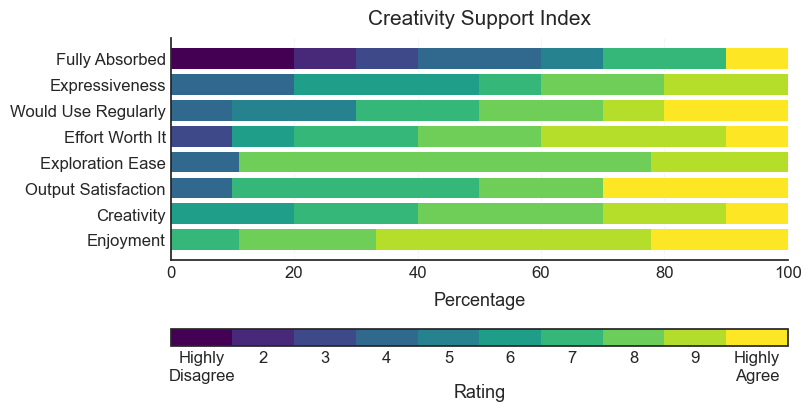}
        \caption{Responses to a questionnaire we adapted from the Creativity Support Index align with our interviews: participants enjoyed using the system and were largely satisfied with its output. However, they mostly didn't report reaching such a level of absorption that they ``forgot about the system or tool'' \cite{CSI}.}
        \label{fig:csi}
        \Description{Creativity support index results. On a scale of 1 (highly disagree) to 10 (highly agree), fully absorbed has 20 percentage of 1, 10 percentage of 2 and 3, 20 percentage of 4, 10 percentage of 5, 20 percentage of 7 and 10 percentage of 10. Expressiveness has 20 percentage of 4, 30 percentage of 6, 10 percentage of 7, 20 percentage of 8 and 9. Would use regularly has 10 percentage of 4, 20 percentage of 5, 20 percentage of 7 and 8, 10 percentage of 9, and 20 percentage of 10. Effort worth it has 10 percentage of 3 and 6, 20 percentage 7 and 8, 30 percentage of 9, and 10 percentage of 10. Approximately, exploration ease has 10 percentage of 4, 60 percentage of 8 and 30 percentage of 9. Output satisfaction has 10 percentage of 4, 40 percentage of 6, 20 percentage of 2, and 30 percentage of 10. Creativity has 20 percentage of 6, 20 percentage of 7, 30 percentage of 8, 20 percentage of 9, and 10 percentage of 10. Approximately, enjoyment has 10 percentage of 7, 20 percentage of 8, 50 percentage of 9, and 20 percentage of 10.}
    \end{minipage}
    \hfill
    \begin{minipage}{0.48\textwidth}
        \centering
        \includegraphics[width=\linewidth]{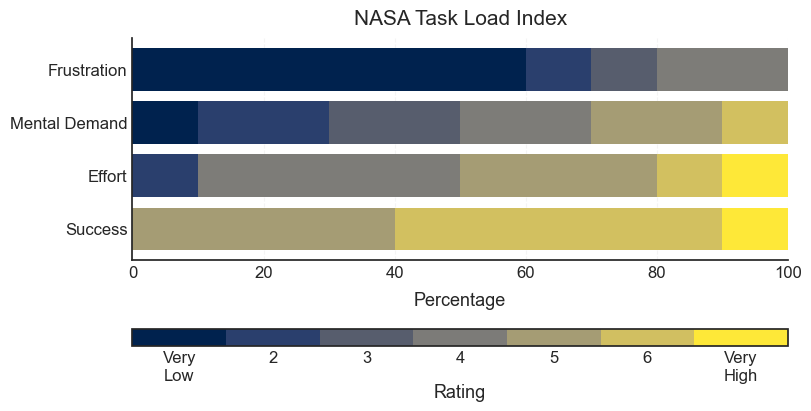}
        \caption{Responses to a questionnaire we adapted from the NASA Task Load index \cite{NASATLX}. Participants mostly responded relatively high to ``How successful were you in accomplishing what you were asked to do?'' and `very low' to ``How insecure, discouraged, irritated, stressed, and annoyed were you?''}
        \label{fig:tlx}
        
        \Description{NASA-TLX results. On the scale of 1 (very low) to 7 (very high), frustration has 60 percentage of 1, 10 percentage of 2, 10 percentage of 3, and 20 percentage of 4. Mental demand has 10 percentage of 1, 20 percentage of 2, 3, 4, and 5, and then 10 percentage of 6. Effort has 10 percentage of 2, 40 percentage of 4, 30 percentage of 5, and 10 percentage of 6 and 7. Success has 40 percentage of 5, 50 percentage of 6, and 10 percentage of 7.}
    \end{minipage}
\end{figure*}

\subsection{Recruitment}

We recruited recently published poets from among a set of poetry journal publications, similar to the methodology of \cite{long_not_2024, booten_poetry_2021} aiming to produce insights along the lines of the case studies of \citet{gabriel_inkwell_2015}. We selected publications based on our knowledge of contemporary poetry practice, for qualities such as their reputation for experimentation and interest in expressive poetry writing. We attempted to contact every poet with work in the most recently published issue from each journal. After reaching out to \raisebox{0.5ex}{\texttildelow}150 recently published poets, \raisebox{0.5ex}{\texttildelow}15 poets responded, and 10 writers completed the full study. Following \citet{jacobs_supporting_2017}, we paid each writer comparably to a commission. Participant backgrounds are given in Appendix \ref{demographics}.

\subsection{Extended User Study} \label{extended-user-study}

Our study design is informed by that of \citet{long_not_2024}, who hypothesized that due to the \textit{novelty effect}, user perceptions of AI systems' usefulness may temper over time. Extended and longitudinal studies lasting longer than a typical 30-60 minute controlled user study seek to address this problem. \citeauthor{long_not_2024} found that users go through a \textit{familiarization phase} (which may last on the order of weeks) in which their understanding of the system's capabilities changed and writers reported increased ownership over text produced with their writing tool.

\subsection{In-Situ Creative Writing}

As we are interested in observing as close as possible, the use of the system within a real \textit{writing context} \cite{lee_design_2024} or task environment \cite{flower_cognitive_1981}, we chose to focus on how \textit{Phraselette} is used within the process of practicing artists, how it affects the production of a work, the qualities of the produced artifact, and its reception. Following \citet{calderwood_how_2018}, we asked the writers to attempt to integrate the system into their writing practice, encouraging them to discover if it could lead to publishable work.

Both as a recruitment incentive and to encourage real-world use, our data policy indicated that we wouldn't use participant data for anything other than reporting study results. Outside of limited excerpts (<1 stanza / paragraph) used for study reporting, we ceded to participants all publishing rights to material they produced with the system.

Writers were given a link to \textit{Phraselette} and invited to join an online Discord server (chatroom) where they might share their work, use-strategies, and support each other with Q\&A style peer mentoring. Ultimately the server did not reach a high level of engagement, beyond an occasional shared poem.

\subsection{Procedure}

We conducted a virtual onboarding for each writer, consisting of a live demonstration of the system and a short interview. Writers were then given free use of the system for two weeks. They were instructed to use the tool for at least 25 minutes on at least five separate days over the course of the study.

We asked each writer to fill out a post-session survey after concluding each day's engagement with the system. The post-session survey contained questions about their intended outcome for the session, the characteristics of the text it produced, and their perceived effect on their writing. 

After completing this task, writers completed for a final interview and a retrospective survey about their perceived success and cognitive load, using the the Creativity Support Index \cite{CSI} and NASA Task Load Index \cite{NASATLX} as reference.

This resulted in two sets of interview transcripts, (interviews guided to address our research questions below). To answer our research questions, we report on the themes that emerged after thematically \cite{braun2006using} coding the interview transcripts, survey results, and analyzing usage logs.

\subsubsection{Interaction Metaphor} \label{mental}

In Section \ref{background_mental_model}, we discussed how an artist's mental model \cite{long_not_2024} or interaction metaphor \cite{lee_design_2024} of a creativity support tool impacts its artmaking affordances. As part of the post-session survey, we asked users to indicate if they used the system as a: \textit{Tool},\textit{ Content Generator}, \textit{Co-pilot}, \textit{Reference Guide}, or \textit{Collaborator}, following \cite{long_not_2024,lee_design_2024}.

\section{Findings}

\subsection{Impressions and Expectations}

\subsubsection{Impressions were positive} 

As reported in Figure \ref{fig:csi}, most writers expressed positivity about the tool after using it for two weeks. P9 stated ``my experience with it overall was very good'', wishing they could use it longer and could envision getting used to ``a dependency growing in a good way''. P4 mentioned that revision felt easier than ``regular writing''. 

P2 was particularly enthusiastic about \textit{Phraselette}. About the language: ``I liked it so much''. His process typically involves ``a regular thesaurus'', but with \textit{Phraselette} , ``it felt like there were more options and just more unusual options, and that was definitely welcome. More is better''. He described a flurry of revisions that came about when he applied \textit{Phraselette} to a stack of older drafts that had been sitting untouched due to various problems. Finding so much success, he turned to other poems that he felt he had previously marked as completed, but now found ways to improve. ``[\textit{Phraselette}] challenges us a lot'', he said.

P6 was the only participant whose overall impression was that the effort put into the tool was not worth it, noting that generation speed could be inconsistent (an issue pointed out by others). ``It seemed extraneous to me'', they said. 

\subsubsection{Play and Fun}

Most participants' first thoughts on the system included some reference to the fun they experienced with the system (P1-P5, P8-P10). Many indicated that its playful nature or their enjoyment of using the tool made them want to use it more than expected. 

\subsubsection{Words as Reward}

P10 describes a moment where she found a phrase that ``made it all worth it''. Like other participants, she described working with the study contract as a seed text; like other participants (P4, P5,) she indicated that early experiments produced text that didn't fit her goals, but the enjoyment of seeing new language appear kept her querying. As a writing teacher, she instructs students about textual provenance and began looking for words related to language, rivers, fluidity, as well as ownership. It wasn't until after a long ``playful'' session that `Phraselette'' was converted to ``praiselet'' and eventually to ``phraseletter'', the word that ``made it all worth it'' for its connotations of `renting' and fluidity. 

P2 tells a similar story, to arrive at the line ``long since I transposed myself into a song'', which mostly came directly from the system and had an outsized effect on their practice.

\subsubsection{Users compared Phraselette to other AI systems}

P10, who had no experience with AI writing tools, reported that despite their initial ``fear and anxiety about AI'', \textit{Phraselette}'s creative control left them feeling more open-minded about AI writing: ``I must say, I'm sort of relieved to be not working in a ChatGPT context, because a lot of my younger colleagues are finding it tough to navigate that''. 

P4 said their initial thoughts were to use it to create ``brand new sentences from nothing'', as is typical with other AI writing tools. This is in contrast to the use they eventually found, which involved taking multiple passes through the text as \textit{Phraselette} produced new images to explore.

P6 mentioned that they hoped that it would produce writing different from other AI systems to produce long form text. They struggled to ``get it to merge two different contexts''. Only when they started to use it for shorter phrases did they report: ``once I figured out the best way to use it, then I was very happy with it.''

On the other hand P7, a poet with a level of experience with AI tools, compared \textit{Phraselette} to a surrealist writing tool they've now lost to the internet, which they described as providing ``random surrealistic phrases''. Though they were dissimilar to the draft context, these compared favorably to \textit{Phraselette}'s rephrasings as ``beautiful fragments of language'' that could stand alone. 

P8 said ``The way that we generally look at AI, it's kind of an abomination, you know, like we view it as kind of a meddling, sort of, you know. And I didn't feel like [\textit{Phraselette}] was meddling, in the way that it made me relinquish ownership of the text, I felt like it was the opposite... freeing... liberating... generative sort of not even in an automatic way, but in a manual way, almost. It still took work to mine the new language. I guess it was kind of like farming or something like that.''

\begin{figure}
    \centering
    \includegraphics[width=1.05\linewidth]{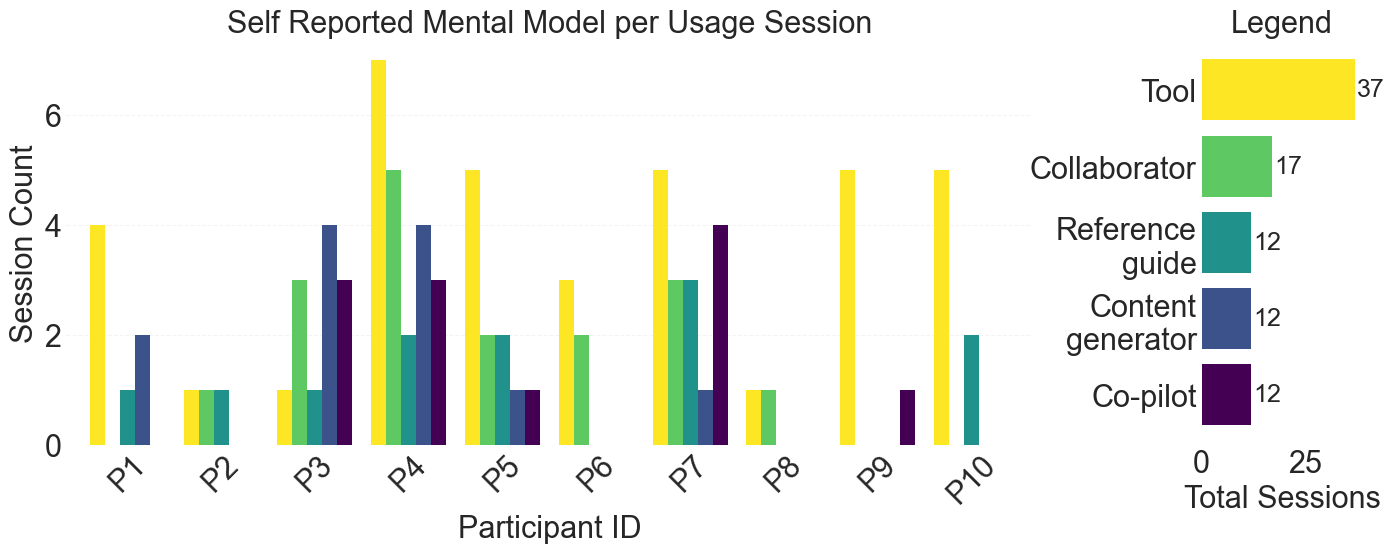}
    \caption{In response to ``I used the system as a \_\_\_\_\_'', writers most often answered `Tool' after each prolonged interaction with Phraselette.}
    \label{fig:mental-model-results}
    \Description{The right side of this figure shows "Self Reported Mental Model per Usage Session" for each participant. P1 has four tools, one reference guide, and two content generators. P2 has one count for tool, collaborator, and reference guide. P3 has one for tool, three for collaborator, one for reference guide, four for content created and three co-pilot. P4 has seven tool, five collaborator, two reference guide, four content generator, and three co-pilot. P5 has five for tool, two for collaborator and reference guide, and one for content generator and co-pilot. P6 has three for tool and two for collaborator. P7 has five for tool, three for collaborator and reference guide, one for content generator, and four  for co-pilot. P8 has one for tool and collaborator. P9 has five for tool and one for co-pilot. P10 has five for tool and two for reference guide. The left side shows legend and total count for each item: tool has 37, collaborator has 17, and reference guide, content generator, and co-pilot each has 12.}
\end{figure}

\subsection{Interaction Metaphor}

Responses to the post-session survey indicate that writers most often described the system as a tool (Figure \ref{fig:mental-model-results}). This may align with our findings in Section \ref{findings-ownership}, in which users mostly reported feeling ownership over the text they produced with the system. That is, we expect a sense of ownership to be tied to the perception of a system as a tool, rather than as, e.g., a collaborator or co-pilot.

Many participants continued to circle back on notions of the system's mental model or interaction metaphor throughout the interviews. Two participates (P2, P10) brought up their liking of the co-pilot metaphor, and P2 noted that this view is not incompatible with a sense of ownership.

\subsection{What Writers Did}

\subsubsection{Phraselette was used to Revise} Writers reported the most success when they used it to revise. Many (including P1-P4, P7) came to the study with existing projects that they reported \textit{Phraselette} having some impact on.

\subsubsection{Phraselette was used for Conceptual Discovery} Many participants (including P3,P4,P6) reported using the tool to bring them to new concepts and images. P3 mentioned using it to directly revise as well as to guide their writing. P6 described their process as an accordion, looking for ``big explosive moments'' before editing these down. P4 said that``I felt myself going in directions that I normally wouldn't, in terms of, the various landscapes it created.'' They mentioned that the context tool was prone to `traveling' to new images. While working on a naturalistic poem, it funneled their attention such that ``instead of going leaf to tree to grass'' (a series of short conceptual leaps) it sent them from ``leaf to tree to boat... a bigger leap.. and now we're on the sea, we're traveling to an island.''

\subsubsection{Writers used Multiple Wells In Conjunction}

All writers used more than one Well. Many (including P2, P3, P4) described workflows that ran different Well types simultaneously. The thesaurus and reader Wells were most often mentioned as providing some value in the writers' eventual workflow. P2 and P4 differed from other writers in that they mentioned the Context Well as one they engaged in most, though both also mentioned combining it with the reader or thesaurus. They discussed finding value in shorter phrase lengths (using the word length constraint). 

Some (including P2, P3) reported mostly using the context Well in conjunction with the reader. P3 said ``I was, like, popping on these different visual filters on a phrase and being like, okay, that's kind of cool. That's like how a 19th century poet might phrase this particular line, okay, this is how a medievalist might rephrase.'' He mentioned viewing the system as a series of lenses or prisms---our original intended interaction metaphor for Phraselette, which we used to frame the system internally during the early phases of development.

\subsubsection{`Misreads' as Alternate Interpretations} \label{misreads}

Participants (including P4, P7, P8, P10) noted a utility in finding a difference between system's interpretation and their own (Section \ref{workshop_material}).

P4 noted that ``[the context well] oftentimes had small errors or misreads [...] And that's what was interesting to me, right? Because this is when we're, like, moving more into, like, the poetic space''. Other poets, including P10, brought up the misread as the most interesting use case of the system (perhaps connecting to `lost poetry' in Section \ref{lost}). 

\subsubsection{Translations} \label{translation}

P8, a translator of poetry, said that it ``kind of opened my mind, if that makes sense... it answers a new dimension in the production of [a translation] and kind of just makes it come alive. It makes it feel a lot more dynamic, and makes the parts feel a lot more movable, sort of more like a puzzle piece, which is challenge that I [face in translation] where a certain phrase or colloquialism or something that might be regarded as untranslatable will seem so essential.''

\subsubsection{Workshopping}

Participants (including P5, P7) commented that the Reader Well's feedback felt appropriate and useful. P5 was ``impressed'' that the critical feedback provided by the reader was able to discern that an image (a pool of melted flesh and a scorched wick) was meant to represent a sense of shame and diminished self worth, and started thinking of \textit{Phraselette} as ``a feedback tool''.

The feedback ``reminded [P5] of a [writing] workshop''. She reported that the reader Well could be useful in order to hear contrasting voices---e.g., when she gave it the persona of a skateboarder. ``Like being in workshop with somebody who writes very differently from you and like, doesn't understand what you're doing'', forcing one to take a step back. At other times, it offered praise: ``I really loved it when it, like, praised the poem,'' she said.



\subsubsection{Workshopping provided interpretive material} \label{workshop_material} P7 mirrored the workshop metaphor. They expressed a balanced apprehension in the notion of such access to feedback: ``You don't really understand why you write something, until it's over''. They explained that deciding too soon what a poem is about comes across as didactic. Despite this, they came down positively on the feedback, saying it gets them to ask ``Is that what it's about?'' They speculated that the feedback might make one \textit{resist} the interpretation provided by the system: ``Oh, is it reading about x? Because that's not what I intended it to be, and therefore I should revise.'' (See Discussion).

P8 had a similar, quite positive take about how he used the  interpretive material provided by the system. It let him become, ``not distracted by the connotations of the original thing that I've sort of described or prescribed to the language... it's like a chisel breaking apart those connotations'' (See Discussion).



\subsection{Ownership} \label{findings-ownership}

\subsubsection{Most writers maintained a sense of ownership}

P2, who utilized many of the suggestions \textit{Phraselette} produced, described a ``radical change'' between the original poem and the result they came up with using Phraselette's suggestions. Asked if they still felt ownership over the text, they responded ``Yes, I did... although expressions have changed, words have changed the syntax has changed, the ideas are still original'' (their own). Elsewhere, they stated ``I wrote the poem and I know what is, what is expected to be there, what's supposed to be there. So there was no change in ownership.'' 

P3 felt that using \textit{Phraselette} was ``no different than using a thesaurus'', noting that ``found text poetry is a totally legitimate technique'', and that he was ``not too precious'' when it comes to taking text from other sources: ``obviously I don't plagiarize, but I did feel ownership of the text''. 

P6 mentioned that, though they sometimes use Grammarly to rewrite text to ``sound like overly upbeat marketing,`` \textit{Phraselette}, generating only single words, didn't make them question ``Am I cheating a little bit in using this?''---a feeling they implied might arise from the former tool---but rather that ``I felt ownership over single words.''

P9 ``rarely changed things'' from their starting text, which they attributed to their own sense of control: ``I didn't realize the extent to which I really committed to what I put down in the first place''.

\subsubsection{Some writers noted an alternate ownership model}

P10 relates that ``ownership is a funny word'' in conversation with their neologism ``phraseletter'' (Section \ref{findings-ownership}). P4 said that individual rephrasings ``did not feel mine'' in the same way, ``selecting or curating from the responses of the machine... does feel somewhat human, but not to the same degree, right? It felt like I was just the curator or editor almost''. P8, asked if they felt ownership over the text, said ``I'm kind of an automatic writer and subscribe to the dada thing... the idea that language sort of happens to us.'' This challenges the frequently assumed value of 'ownership' in writing practices: not all writing cultures (see, e.g., Section \ref{material_writing_support}) center ownership in their definition of what it means to be a successful writer.

\subsubsection{Disclosure of AI text}

Most poets indicated that they planned to share the produced writing without emphasizing or even disclosing their use of the AI system. This finding (which surprised the interviewer/first author) is put forward as an indication that the mental model writers adopted for the system is more comparable to other digital tools (such as a text editor) or thesaurus, rather than autonomous writing systems like ChatGPT, for which the prevailing sentiment is one of informed disclosure. This may complicate understanding of the \textit{AI ghostwriter effect}, in which writers do not consider themselves the authors of AI-generated text but do not declare AI authorship \cite{draxler_ai_2024}.

\subsection{Rephrasings and Style}

\subsubsection{Rephrasings were expansive} Many participants noted the sometimes surprising ability of the system to write in accordance with a given writer or style. P9 used it to mirror Louise Glück: ``I'm seeing how closely this is feeling to her exact verbiage. I mean, it's kind of amazing.'' 

P2 returned to the term ``all embracing'' to describe the expansive space of language that was produced by \textit{Phraselette}---focusing on the breadth of options it provided not ``sticking to one profile''. They said \textit{Phraselette} allowed them to dive deeper than ``regular'' writers, using it to produce metaphors of loneliness using the Justice League character J'onn J'onzz and transliterated drama of Nobel Laureate Wole Soyinka.

P5 found rephrasings to be ``beautiful'' and``vivid'', using the tool to strike the tone of a passionist priest (a Christian congregation focused on the crucifixion). They described ``comprehensive'' and ``creative'' multi-word revisions: you couldn't ``go on a site and find that with a traditional thesaurus.'' P4 was described that rephrasings ``felt good'', ``well crafted'', and also ``a little odd'', ``dense''.


\subsubsection{Uninteresting text was useful when it was targeted}

P8 felt that the system's text style was generally ``not good'' and ``totally crazy'', but was enthusiastic about text that hit a narrow mark necessary to satisfy a translation.


P7 described rephrasings as ``fun, funky, odd, quirky, unusual, eccentric''. They reported their first sessions as unsuccessful: they attempted to integrate \textit{Phraselette} into their practice of ``poetry of the mundane, poetry of the quotidian'' and found the rephrasings clunky and out of place, noting more value in the reader feedback. Like some other participants (including P3, P9, and P10), P7 didn't realize until late in the study that they could enter their own descriptions of readers; when they did, ``it became really powerful'' and allowed them to write with the ``focus'' of a given character.

\subsubsection{Writer voice was maintained}

P3 says `` I don't mean this in a negative way. It did feel just like my general writing... I didn't necessarily feel like I was producing work that was mind-blowingly unique to me, like it felt like I was just writing poems, which is good. At the same time, I don't think I would have written those poems without the tool.'' This may be an instance of  ``inspiration through observation,'' in which looking at AI-generated suggestions shapes what gets written---even when the user doesn't adopt the AI text \cite{roemmele2021inspiration}. P8 agreed: ``[the suggestions] felt familiar... I tended to use tools that agreed with me''. 




\subsection{Rephrasings as Material} \label{findings_material}

P7 described the practice of ``run[ning] through a lot of generations to find something useful'' as an exercise in \textit{curation}. This mirrored P4 describing their role as ``curator or editor''.

P9 described pulling a handful of words from a book of poetry as a way to ``keep me honest, as I tend to drift into abstraction.'' With \textit{Phraselette}, ``a word bank is being presented to us that we didn't come up with on our own, but [which] definitely relates to the subject matter. It's an opportunity to enlarge the poem in one way or another.... viewing these as rephrasings might not be the right way to think about this. Maybe they should be going into this \textit{collective word bank} that, just like floats around and can grow.'' P2 called these words a ``collection'' and ``smorgasbord''.

\subsection{Other Findings}

\subsubsection{Potential for Distraction} \label{distraction}

A few of the participants (P10, P4) speculated that others may find it distracting, connecting this to both their enjoyment of the system and to the number of options it provided. P10 wondered if the ``rabbit hole'' or ``dice roll'' might get in the way of the concentration that desired by some writers: ``What's going to turn up now, what's going to turn up now?'' P8 said they initially ``got too lost in just playing with the tool, which was fun'' before they learned to use it for narrow translation tasks. 

\subsubsection{Publication Quality Poetry}

One way that we might ascertain the success of a poetry writing system is by asking if the poetry is at the level that might be shared or published. 

During recruitment and onboarding, we asked poets to use \textit{Phraselette} in their public artmaking practice. At the end of the study, we asked nearly all writers (skipping P5 and P6 because we ran out of time) if the writing they produced with \textit{Phraselette} would make it into the world. All poets who responded indicated that they would be submitting some work written with \textit{Phraselette} for publication. Many commented on the tool's effect on the poetry.

More than just direct text, P10 noted ``I have confidence that some of the playfulness and exploration will make it into [their long term project]''. This may be suggestive of indirect system effects on user attitudes, as observed by \citet{Jakesch}.

\subsection{System Feedback}

During the interviews, participants largely refrained from commenting on system limitations that might be viewed as specific to the current version of the implementation, whereas these comments were more pronounced in the post-session survey, especially earlier in the study.

\subsubsection{Assorted Bugs} We expect the presence of certain bugs did limit the tools ability to absorb into the creative process. P8 relates ``I had a great time. From a creative standpoint... A little buggy when I tried to use many palettes, it crashed or loaded infinitely.  I didn't use it extensively for that reason.''

\subsubsection{Sound}

Participants (P2, P5, P9, P10) mentioned a lack of success in using the sound constraint during the interviews, and (P1, P3, and P10) mentioned the same difficulty in the post-session survey. We discuss this in Section \ref{future-constraint}.

\subsubsection{Speed}

(P6, P7) noted that it was ``inconsistent in terms of speed'', a known issue. However, the poets who mostly used the context Well noted it to be quick. 

\section{Discussion}

Participant gravitation toward using \textit{Phraselette} for reflection, workshopping, and reinterpretation may be understood in terms of ``intent elicitation''~\cite{kreminski2024intent}. Poets in our study were actively looking to form new interpretations of texts and have their preconceptions challenged; they looked for and responded to instances where the system exhibited signs of \textit{interpretation} as well as \textit{intent} and found value in pushing against these, as well as in realigning their texts towards alternate intents and interpretations. These findings connect to a range of prior work on how writing tools can move beyond manipulating a working draft to engage the writer's sensorium and shape interpretive processes~\cite{buschek_collage_2024,singh_where_2022,gero_metaphoria_2019,kreminski_reflective_nodate,booten_lotus_2023}, as well as how writing tools can support writers without writing for them \cite{WithoutWriting1,WithoutWriting2}.



Section \ref{translation} and \ref{workshop_material} highlight a design bridge between Buschek's discussion of the value of systematic `fragmentation' of texts \cite{buschek_collage_2024} and Kreminski's view of the affordances of computational search in poetry writing \cite{kreminski_computational_2024}. One user's description of \textit{Phraselette} as a conceptual ``chisel'' suggests a view of the system as a defamiliarizing force \cite{calderwood_how_2018}, a ``loser of poetry'' and its meaning.

Our findings around user mental models of \textit{Phraselette} reinforce both \citeauthor{li_beyond_2023}'s theory of ``normative ground''~\cite{li_beyond_2023} and \citeauthor{Reframer}'s findings regarding when users perceive co-creative systems as collaborators rather than tools~\cite{Reframer}. In particular, \textit{Phraselette}'s placing of initiative firmly within the hands of the user may contribute to broad user agreement that the system is best viewed as a tool.






\section{Limitations and Future Work}

\label{future-constraint}

\label{future-constraint}

Participants struggled to make use of the Sound constraint and revising it is a clear next step. As is fixing a bug in which readers sometimes switched `inexplicably' between French, Spanish, and English.



\textit{Phraselette} was designed as an extensible platform to support a diverse array of constraints and generators (Section \ref{extensible}). The affordances gained from composing generation strategies and different forms of constraint gain more utility if constraints are aligned with writers' understanding. 
Naturally, we will continue introducing new constraints and generators to this framework, including a BabelNet~\cite{BabelNet} graph search algorithm and a word embedding model that can measure and guide phrase generation. 

We are convinced by the argument \cite{li_beyond_2023} provides that reducing the rigidity of the abstractions mandated by the system is a key factor that describes how the \emph{vertical movement} of creative tools' can transfer power from the tool designer to the artist. Phraselette's notion of constraint---the part of speech and sound constraints specifically---imply a set of values that are not shared by all poets. Computational search along boundaries that can be well understood on both sides of the human/machine divide is a design challenge with implications extending into the future of poetic practice---the boundaries \textit{and misrepresentations} that constraints afford lend a voice to the instrument.

\section{Conclusion}

As an alternative to the normative ground of automated writing tools like ChatGPT, we have characterized a design ethos---\emph{material writing support}---that we argue to be better aligned with the \emph{writerly values} of experimental poets. A system designed in line with this ethos, \textit{Phraselette}, holds up well in an extended expert evaluation with 10 published poets, showcasing both the potential of design to shift user mental models of text generation technologies and their usage, and the value of machine-suggested short phrases and textual reinterpretations to help poets arrive at interpretations and intents within the drafting process. Our work provides both a case study of designing writing support systems to adhere to writerly values and a template for the design of future systems in a similar mold.

\bibliography{paper}


\begin{thebibliography}{66}


\ifx \showCODEN    \undefined \def \showCODEN     #1{\unskip}     \fi
\ifx \showDOI      \undefined \def \showDOI       #1{#1}\fi
\ifx \showISBNx    \undefined \def \showISBNx     #1{\unskip}     \fi
\ifx \showISBNxiii \undefined \def \showISBNxiii  #1{\unskip}     \fi
\ifx \showISSN     \undefined \def \showISSN      #1{\unskip}     \fi
\ifx \showLCCN     \undefined \def \showLCCN      #1{\unskip}     \fi
\ifx \shownote     \undefined \def \shownote      #1{#1}          \fi
\ifx \showarticletitle \undefined \def \showarticletitle #1{#1}   \fi
\ifx \showURL      \undefined \def \showURL       {\relax}        \fi
\providecommand\bibfield[2]{#2}
\providecommand\bibinfo[2]{#2}
\providecommand\natexlab[1]{#1}
\providecommand\showeprint[2][]{arXiv:#2}

\bibitem[Agarwal et~al\mbox{.}(2024)]%
        {TowardWesternStyles}
\bibfield{author}{\bibinfo{person}{Dhruv Agarwal}, \bibinfo{person}{Mor Naaman}, {and} \bibinfo{person}{Aditya Vashistha}.} \bibinfo{year}{2024}\natexlab{}.
\newblock \showarticletitle{{AI} suggestions homogenize writing toward western styles and diminish cultural nuances}.
\newblock \bibinfo{journal}{\emph{arXiv preprint arXiv:2409.11360}} (\bibinfo{year}{2024}).
\newblock


\bibitem[Anderson et~al\mbox{.}(2024)]%
        {HomogenizationEffects}
\bibfield{author}{\bibinfo{person}{Barrett~R Anderson}, \bibinfo{person}{Jash~Hemant Shah}, {and} \bibinfo{person}{Max Kreminski}.} \bibinfo{year}{2024}\natexlab{}.
\newblock \showarticletitle{Homogenization effects of large language models on human creative ideation}. In \bibinfo{booktitle}{\emph{Proceedings of the 16th Conference on Creativity \& Cognition}}. \bibinfo{pages}{413--425}.
\newblock


\bibitem[Arnold et~al\mbox{.}(2020)]%
        {PredictableWriting}
\bibfield{author}{\bibinfo{person}{Kenneth~C Arnold}, \bibinfo{person}{Krysta Chauncey}, {and} \bibinfo{person}{Krzysztof~Z Gajos}.} \bibinfo{year}{2020}\natexlab{}.
\newblock \showarticletitle{Predictive text encourages predictable writing}. In \bibinfo{booktitle}{\emph{Proceedings of the 25th International Conference on Intelligent User Interfaces}}. \bibinfo{pages}{128--138}.
\newblock


\bibitem[Arnold et~al\mbox{.}(2021)]%
        {WithoutWriting1}
\bibfield{author}{\bibinfo{person}{Kenneth~C Arnold}, \bibinfo{person}{April~M Volzer}, {and} \bibinfo{person}{Noah~G Madrid}.} \bibinfo{year}{2021}\natexlab{}.
\newblock \showarticletitle{Generative Models can Help Writers without Writing for Them}. In \bibinfo{booktitle}{\emph{IUI Workshops}}.
\newblock


\bibitem[Beiguelman(2006)]%
        {nomadic_poetry}
\bibfield{author}{\bibinfo{person}{Giselle Beiguelman}.} \bibinfo{year}{2006}\natexlab{}.
\newblock \showarticletitle{Nomadic Poetry}.
\newblock In \bibinfo{booktitle}{\emph{New Media Poetics: Contexts, Technotexts, and Theories}}. \bibinfo{publisher}{MIT Press}.
\newblock


\bibitem[Booten(2023)]%
        {booten_lotus_2023}
\bibfield{author}{\bibinfo{person}{Kyle Booten}.} \bibinfo{year}{2023}\natexlab{}.
\newblock \showarticletitle{Lotus {Chorus} {Workshop}: {Designing} for {Cognitive} {Overload}}. In \bibinfo{booktitle}{\emph{Proceedings of the Eleventh Conference on Computation, Communication, Aesthetics \& X}}. \bibinfo{publisher}{i2ADS — Research Institute in Art, Design and Society}.
\newblock
\urldef\tempurl%
\url{https://doi.org/10.34626/XCOAX.2023.11TH.185}
\showDOI{\tempurl}


\bibitem[Booten and Gero(2021)]%
        {booten_poetry_2021}
\bibfield{author}{\bibinfo{person}{Kyle Booten} {and} \bibinfo{person}{Katy~Ilonka Gero}.} \bibinfo{year}{2021}\natexlab{}.
\newblock \showarticletitle{Poetry {Machines}: {Eliciting} {Designs} for {Interactive} {Writing} {Tools} from {Poets}}. In \bibinfo{booktitle}{\emph{Creativity and {Cognition}}}. \bibinfo{publisher}{ACM}.
\newblock
\showISBNx{978-1-4503-8376-9}
\urldef\tempurl%
\url{https://doi.org/10.1145/3450741.3466813}
\showDOI{\tempurl}


\bibitem[Braun and Clarke(2006)]%
        {braun2006using}
\bibfield{author}{\bibinfo{person}{Virginia Braun} {and} \bibinfo{person}{Victoria Clarke}.} \bibinfo{year}{2006}\natexlab{}.
\newblock \showarticletitle{Using thematic analysis in psychology}.
\newblock \bibinfo{journal}{\emph{Qualitative research in psychology}} \bibinfo{volume}{3}, \bibinfo{number}{2} (\bibinfo{year}{2006}), \bibinfo{pages}{77--101}.
\newblock


\bibitem[Buschek(2024)]%
        {buschek_collage_2024}
\bibfield{author}{\bibinfo{person}{Daniel Buschek}.} \bibinfo{year}{2024}\natexlab{}.
\newblock \showarticletitle{Collage is the New Writing: Exploring the Fragmentation of Text and User Interfaces in AI Tools}. In \bibinfo{booktitle}{\emph{Proceedings of the 2024 ACM Designing Interactive Systems Conference}}. \bibinfo{pages}{2719--2737}.
\newblock


\bibitem[Calderwood et~al\mbox{.}(2018)]%
        {calderwood_how_2018}
\bibfield{author}{\bibinfo{person}{Alex Calderwood}, \bibinfo{person}{Vivian Qiu}, \bibinfo{person}{Katy~Ilonka Gero}, {and} \bibinfo{person}{Lydia~B Chilton}.} \bibinfo{year}{2018}\natexlab{}.
\newblock \showarticletitle{How {Novelists} {Use} {Generative} {Language} {Models}: {An} {Exploratory} {User} {Study}}. In \bibinfo{booktitle}{\emph{23rd {International} {Conference} on {Intelligent} {User} {Interfaces}}}. \bibinfo{publisher}{ACM}.
\newblock
\showISBNx{978-1-4503-4945-1}


\bibitem[Cayley(2006)]%
        {cayley_time_2006}
\bibfield{author}{\bibinfo{person}{John Cayley}.} \bibinfo{year}{2006}\natexlab{}.
\newblock \showarticletitle{Time {Code} {Language}: {New} {Media} {Poetics} and {Programmed} {Signification}}.
\newblock In \bibinfo{booktitle}{\emph{New {Media} {Poetics} {Contexts}}}.
\newblock
\urldef\tempurl%
\url{https://direct.mit.edu/books/book/2676/chapter-abstract/72465/Time-Code-Language-New-Media-Poetics-and?redirectedFrom=PDF}
\showURL{%
\tempurl}


\bibitem[Chakrabarty et~al\mbox{.}(2024)]%
        {AIWritingSalvaged}
\bibfield{author}{\bibinfo{person}{Tuhin Chakrabarty}, \bibinfo{person}{Philippe Laban}, {and} \bibinfo{person}{Chien-Sheng Wu}.} \bibinfo{year}{2024}\natexlab{}.
\newblock \showarticletitle{Can {AI} writing be salvaged? Mitigating Idiosyncrasies and Improving Human-{AI} Alignment in the Writing Process through Edits}.
\newblock \bibinfo{journal}{\emph{arXiv preprint arXiv:2409.14509}} (\bibinfo{year}{2024}).
\newblock


\bibitem[Cherry and Latulipe(2014)]%
        {CSI}
\bibfield{author}{\bibinfo{person}{Erin Cherry} {and} \bibinfo{person}{Celine Latulipe}.} \bibinfo{year}{2014}\natexlab{}.
\newblock \showarticletitle{Quantifying the creativity support of digital tools through the {Creativity Support Index}}.
\newblock \bibinfo{journal}{\emph{ACM Transactions on Computer-Human Interaction (TOCHI)}} \bibinfo{volume}{21}, \bibinfo{number}{4} (\bibinfo{year}{2014}).
\newblock


\bibitem[Coenen et~al\mbox{.}(2021)]%
        {coenen_wordcraft_2021}
\bibfield{author}{\bibinfo{person}{Andy Coenen}, \bibinfo{person}{Luke Davis}, \bibinfo{person}{Daphne Ippolito}, \bibinfo{person}{Emily Reif}, {and} \bibinfo{person}{Ann Yuan}.} \bibinfo{year}{2021}\natexlab{}.
\newblock \showarticletitle{Wordcraft: a human-{AI} collaborative editor for story writing}.
\newblock \bibinfo{journal}{\emph{arXiv preprint arXiv:2107.07430}} (\bibinfo{year}{2021}).
\newblock


\bibitem[Doherty(2022)]%
        {doherty_kathy_2022}
\bibfield{author}{\bibinfo{person}{Maggie Doherty}.} \bibinfo{year}{2022}\natexlab{}.
\newblock \showarticletitle{Kathy {Acker}’s {Art} of {Identity} {Theft}}.
\newblock \bibinfo{journal}{\emph{The New Yorker}} (\bibinfo{date}{Nov.} \bibinfo{year}{2022}).
\newblock
\showISSN{0028-792X}
\urldef\tempurl%
\url{https://www.newyorker.com/magazine/2022/12/05/kathy-ackers-art-of-identity-theft}
\showURL{%
\tempurl}
\newblock
\shownote{Section: books}.


\bibitem[Doshi and Hauser(2024)]%
        {DoshiHauser}
\bibfield{author}{\bibinfo{person}{Anil~R Doshi} {and} \bibinfo{person}{Oliver~P Hauser}.} \bibinfo{year}{2024}\natexlab{}.
\newblock \showarticletitle{Generative {AI} enhances individual creativity but reduces the collective diversity of novel content}.
\newblock \bibinfo{journal}{\emph{Science Advances}} \bibinfo{volume}{10}, \bibinfo{number}{28} (\bibinfo{year}{2024}), \bibinfo{pages}{eadn5290}.
\newblock


\bibitem[Douglass(2023a)]%
        {usc_future_of_writing_symposium_jeremy_2023}
\bibfield{author}{\bibinfo{person}{Jeremy Douglass}.} \bibinfo{year}{2023}\natexlab{a}.
\newblock \bibinfo{title}{{Writing} {To} and {From} {Language} {Models}}.
\newblock \bibinfo{howpublished}{{USC Future of Writing Symposium}}.
\newblock
\urldef\tempurl%
\url{https://www.youtube.com/watch?v=DivsydaBNaQ}
\showURL{%
\tempurl}


\bibitem[Douglass(2023b)]%
        {douglass_writing_2023}
\bibfield{author}{\bibinfo{person}{Jeremy Douglass}.} \bibinfo{year}{2023}\natexlab{b}.
\newblock \bibinfo{title}{Writing with {AI}--{Large} {Language} {Models} ({GPT}, {ChatGPT} et al)}.
\newblock
\newblock
\urldef\tempurl%
\url{https://docs.google.com/document/d/1wjq8xagV1ugZuONTc4nlERyuTv3e9Bq86XpsYQDkiIY/edit?usp=embed_facebook}
\showURL{%
\tempurl}


\bibitem[Draxler et~al\mbox{.}(2024)]%
        {draxler_ai_2024}
\bibfield{author}{\bibinfo{person}{Fiona Draxler}, \bibinfo{person}{Anna Werner}, \bibinfo{person}{Florian Lehmann}, \bibinfo{person}{Matthias Hoppe}, \bibinfo{person}{Albrecht Schmidt}, \bibinfo{person}{Daniel Buschek}, {and} \bibinfo{person}{Robin Welsch}.} \bibinfo{year}{2024}\natexlab{}.
\newblock \showarticletitle{The {AI} {Ghostwriter} {Effect}: {When} {Users} do not {Perceive} {Ownership} of {AI}-{Generated} {Text} but {Self}-{Declare} as {Authors}}.
\newblock \bibinfo{journal}{\emph{ACM Transactions on Computer-Human Interaction}} \bibinfo{volume}{31}, \bibinfo{number}{2} (\bibinfo{date}{April} \bibinfo{year}{2024}), \bibinfo{pages}{1--40}.
\newblock
\showISSN{1073-0516, 1557-7325}
\urldef\tempurl%
\url{https://doi.org/10.1145/3637875}
\showDOI{\tempurl}


\bibitem[Flower and Hayes(1981)]%
        {flower_cognitive_1981}
\bibfield{author}{\bibinfo{person}{Linda Flower} {and} \bibinfo{person}{John~R. Hayes}.} \bibinfo{year}{1981}\natexlab{}.
\newblock \showarticletitle{A {Cognitive} {Process} {Theory} of {Writing}}.
\newblock \bibinfo{journal}{\emph{College Composition and Communication}} \bibinfo{volume}{32}, \bibinfo{number}{4} (\bibinfo{date}{Dec.} \bibinfo{year}{1981}), \bibinfo{pages}{365}.
\newblock
\showISSN{0010096X}
\urldef\tempurl%
\url{https://doi.org/10.2307/356600}
\showDOI{\tempurl}


\bibitem[Freitag and Al-Onaizan(2017)]%
        {freitag_beam_2017}
\bibfield{author}{\bibinfo{person}{Markus Freitag} {and} \bibinfo{person}{Yaser Al-Onaizan}.} \bibinfo{year}{2017}\natexlab{}.
\newblock \showarticletitle{Beam {Search} {Strategies} for {Neural} {Machine} {Translation}}. In \bibinfo{booktitle}{\emph{Proceedings of the {First} {Workshop} on {Neural} {Machine} {Translation}}}. \bibinfo{pages}{56--60}.
\newblock
\urldef\tempurl%
\url{https://doi.org/10.18653/v1/W17-3207}
\showDOI{\tempurl}
\newblock
\shownote{arXiv:1702.01806 [cs]}.


\bibitem[Fussell(1979)]%
        {fussell_poetic_1979}
\bibfield{author}{\bibinfo{person}{Paul Fussell}.} \bibinfo{year}{1979}\natexlab{}.
\newblock \showarticletitle{Poetic {Meter} and {Poetic} {Form}}.
\newblock \bibinfo{journal}{\emph{University of Pennsylvania}} (\bibinfo{year}{1979}).
\newblock


\bibitem[Gabriel et~al\mbox{.}(2015)]%
        {gabriel_inkwell_2015}
\bibfield{author}{\bibinfo{person}{Richard~P. Gabriel}, \bibinfo{person}{Jilin Chen}, {and} \bibinfo{person}{Jeffrey Nichols}.} \bibinfo{year}{2015}\natexlab{}.
\newblock \showarticletitle{{InkWell}: {A} {Creative} {Writer}'s {Creative} {Assistant}}. In \bibinfo{booktitle}{\emph{Proceedings of the 2015 {ACM} {SIGCHI} {Conference} on {Creativity} and {Cognition}}}. \bibinfo{publisher}{ACM}, \bibinfo{address}{Glasgow United Kingdom}, \bibinfo{pages}{93--102}.
\newblock
\showISBNx{978-1-4503-3598-0}
\urldef\tempurl%
\url{https://doi.org/10.1145/2757226.2757229}
\showDOI{\tempurl}


\bibitem[Gero and Chilton(2019a)]%
        {gero_how_2019}
\bibfield{author}{\bibinfo{person}{Katy~Ilonka Gero} {and} \bibinfo{person}{Lydia~B. Chilton}.} \bibinfo{year}{2019}\natexlab{a}.
\newblock \showarticletitle{How a {Stylistic}, {Machine}-{Generated} {Thesaurus} {Impacts} a {Writer}'s {Process}}. In \bibinfo{booktitle}{\emph{Proceedings of the 2019 Conference on {Creativity} and {Cognition}}}. \bibinfo{publisher}{ACM}, \bibinfo{address}{San Diego CA USA}, \bibinfo{pages}{597--603}.
\newblock
\showISBNx{978-1-4503-5917-7}
\urldef\tempurl%
\url{https://doi.org/10.1145/3325480.3326573}
\showDOI{\tempurl}


\bibitem[Gero and Chilton(2019b)]%
        {gero_metaphoria_2019}
\bibfield{author}{\bibinfo{person}{Katy~Ilonka Gero} {and} \bibinfo{person}{Lydia~B. Chilton}.} \bibinfo{year}{2019}\natexlab{b}.
\newblock \showarticletitle{Metaphoria: {An} {Algorithmic} {Companion} for {Metaphor} {Creation}}. In \bibinfo{booktitle}{\emph{Proceedings of the 2019 {CHI} {Conference} on {Human} {Factors} in {Computing} {Systems}}}. \bibinfo{publisher}{ACM}, \bibinfo{address}{Glasgow Scotland Uk}, \bibinfo{pages}{1--12}.
\newblock
\showISBNx{978-1-4503-5970-2}
\urldef\tempurl%
\url{https://doi.org/10.1145/3290605.3300526}
\showDOI{\tempurl}


\bibitem[Gupta et~al\mbox{.}(2023)]%
        {BiasRunsDeep}
\bibfield{author}{\bibinfo{person}{Shashank Gupta}, \bibinfo{person}{Vaishnavi Shrivastava}, \bibinfo{person}{Ameet Deshpande}, \bibinfo{person}{Ashwin Kalyan}, \bibinfo{person}{Peter Clark}, \bibinfo{person}{Ashish Sabharwal}, {and} \bibinfo{person}{Tushar Khot}.} \bibinfo{year}{2023}\natexlab{}.
\newblock \showarticletitle{Bias runs deep: Implicit reasoning biases in persona-assigned {LLMs}}.
\newblock \bibinfo{journal}{\emph{arXiv preprint arXiv:2311.04892}} (\bibinfo{year}{2023}).
\newblock


\bibitem[Hart and Staveland(1988)]%
        {NASATLX}
\bibfield{author}{\bibinfo{person}{Sandra~G Hart} {and} \bibinfo{person}{Lowell~E Staveland}.} \bibinfo{year}{1988}\natexlab{}.
\newblock \showarticletitle{Development of NASA-TLX (Task Load Index): Results of empirical and theoretical research}.
\newblock In \bibinfo{booktitle}{\emph{Advances in Psychology}}. Vol.~\bibinfo{volume}{52}. \bibinfo{publisher}{Elsevier}, \bibinfo{pages}{139--183}.
\newblock


\bibitem[Hayles(2006)]%
        {morris_time_2006}
\bibfield{author}{\bibinfo{person}{N.~Katherine Hayles}.} \bibinfo{year}{2006}\natexlab{}.
\newblock \showarticletitle{The {Time} of {Digital} {Poetry}: {From} {Object} to {Event}}.
\newblock In \bibinfo{booktitle}{\emph{New {Media} {Poetics}}}, \bibfield{editor}{\bibinfo{person}{Adalaide Morris} {and} \bibinfo{person}{Thomas Swiss}} (Eds.). \bibinfo{publisher}{The MIT Press}, \bibinfo{pages}{181--210}.
\newblock
\showISBNx{978-0-262-28021-1}
\urldef\tempurl%
\url{https://doi.org/10.7551/mitpress/5002.003.0013}
\showDOI{\tempurl}


\bibitem[Hermansson and Saar(2017)]%
        {hermansson_nomadic_2017}
\bibfield{author}{\bibinfo{person}{Carina Hermansson} {and} \bibinfo{person}{Tomas Saar}.} \bibinfo{year}{2017}\natexlab{}.
\newblock \showarticletitle{Nomadic writing $^{\textrm{1}}$ in early childhood education}.
\newblock \bibinfo{journal}{\emph{Journal of Early Childhood Literacy}} \bibinfo{volume}{17}, \bibinfo{number}{3} (\bibinfo{date}{Sept.} \bibinfo{year}{2017}), \bibinfo{pages}{426--443}.
\newblock
\showISSN{1468-7984, 1741-2919}
\urldef\tempurl%
\url{https://doi.org/10.1177/1468798417712341}
\showDOI{\tempurl}


\bibitem[Holdeman(2023)]%
        {holdeman_introduction_nodate}
\bibfield{author}{\bibinfo{person}{David Holdeman}.} \bibinfo{year}{2023}\natexlab{}.
\newblock \showarticletitle{Introduction: {Yeats} and {Materiality}}.
\newblock \bibinfo{journal}{\emph{International Yeats Studies}} \bibinfo{volume}{7}, \bibinfo{number}{1} (\bibinfo{year}{2023}).
\newblock


\bibitem[Howe(2013)]%
        {howe_reading_nodate}
\bibfield{author}{\bibinfo{person}{Daniel~C Howe}.} \bibinfo{year}{2013}\natexlab{}.
\newblock \showarticletitle{Reading, writing, resisting: literary appropriation in \textit{The Readers Project}}. In \bibinfo{booktitle}{\emph{Proceedings of the 19th International Symposium on Electronic Art}}. \bibinfo{pages}{178--181}.
\newblock


\bibitem[Jacobs et~al\mbox{.}(2017)]%
        {jacobs_supporting_2017}
\bibfield{author}{\bibinfo{person}{Jennifer Jacobs}, \bibinfo{person}{Sumit Gogia}, \bibinfo{person}{Radomír Mĕch}, {and} \bibinfo{person}{Joel~R. Brandt}.} \bibinfo{year}{2017}\natexlab{}.
\newblock \showarticletitle{Supporting {Expressive} {Procedural} {Art} {Creation} through {Direct} {Manipulation}}. In \bibinfo{booktitle}{\emph{Proceedings of the 2017 {CHI} {Conference} on {Human} {Factors} in {Computing} {Systems}}}. \bibinfo{publisher}{ACM}, \bibinfo{address}{Denver Colorado USA}, \bibinfo{pages}{6330--6341}.
\newblock
\showISBNx{978-1-4503-4655-9}
\urldef\tempurl%
\url{https://doi.org/10.1145/3025453.3025927}
\showDOI{\tempurl}


\bibitem[Jakesch et~al\mbox{.}(2023)]%
        {Jakesch}
\bibfield{author}{\bibinfo{person}{Maurice Jakesch}, \bibinfo{person}{Advait Bhat}, \bibinfo{person}{Daniel Buschek}, \bibinfo{person}{Lior Zalmanson}, {and} \bibinfo{person}{Mor Naaman}.} \bibinfo{year}{2023}\natexlab{}.
\newblock \showarticletitle{Co-writing with opinionated language models affects users’ views}. In \bibinfo{booktitle}{\emph{Proceedings of the 2023 CHI Conference on Human Factors in Computing Systems}}.
\newblock


\bibitem[Katnoria(2020)]%
        {katnoria_visualising_2020}
\bibfield{author}{\bibinfo{person}{Katnoria}.} \bibinfo{year}{2020}\natexlab{}.
\newblock \bibinfo{title}{Visualising {Beam} {Search} and {Other} {Decoding} {Algorithms} for {Natural} {Language} {Generation}}.
\newblock
\newblock
\urldef\tempurl%
\url{https://medium.com/voice-tech-podcast/visualising-beam-search-and-other-decoding-algorithms-for-natural-language-generation-fbba7cba2c5b}
\showURL{%
\tempurl}


\bibitem[Kelly et~al\mbox{.}(2023)]%
        {kelly_there_2023}
\bibfield{author}{\bibinfo{person}{Jack Kelly}, \bibinfo{person}{Alex Calderwood}, \bibinfo{person}{Noah Wardrip-Fruin}, {and} \bibinfo{person}{Michael Mateas}.} \bibinfo{year}{2023}\natexlab{}.
\newblock \showarticletitle{There and {Back} {Again}: {Extracting} {Formal} {Domains} for {Controllable} {Neurosymbolic} {Story} {Authoring}}.
\newblock \bibinfo{journal}{\emph{Proceedings of the AAAI Conference on Artificial Intelligence and Interactive Digital Entertainment}} \bibinfo{volume}{19}, \bibinfo{number}{1} (\bibinfo{date}{Oct.} \bibinfo{year}{2023}), \bibinfo{pages}{64--74}.
\newblock
\showISSN{2334-0924, 2326-909X}
\urldef\tempurl%
\url{https://doi.org/10.1609/aiide.v19i1.27502}
\showDOI{\tempurl}


\bibitem[Kirschenbaum(2001)]%
        {kirschenbaum_materiality_2001}
\bibfield{author}{\bibinfo{person}{Matthew Kirschenbaum}.} \bibinfo{year}{2001}\natexlab{}.
\newblock \showarticletitle{Materiality and {Matter} and {Stuff}: {What} {Electronic} {Texts} {Are} {Made} {Of}}.
\newblock  (\bibinfo{year}{2001}).
\newblock
\urldef\tempurl%
\url{https://electronicbookreview.com/essay/materiality-and-matter-and-stuff-what-electronic-texts-are-made-of/}
\showURL{%
\tempurl}


\bibitem[Kirschenbaum(2016)]%
        {kirschenbaum_track_2016}
\bibfield{author}{\bibinfo{person}{Matthew~G Kirschenbaum}.} \bibinfo{year}{2016}\natexlab{}.
\newblock \bibinfo{booktitle}{\emph{Track changes: {A} literary history of word processing}}.
\newblock \bibinfo{publisher}{Harvard University Press}.
\newblock


\bibitem[Kreminski(2024a)]%
        {kreminski_computational_2024}
\bibfield{author}{\bibinfo{person}{Max Kreminski}.} \bibinfo{year}{2024}\natexlab{a}.
\newblock \showarticletitle{Computational {Poetry} is {Lost} {Poetry}}. In \bibinfo{booktitle}{\emph{Proceedings of the {Halfway} to the {Future} {Symposium}}}. \bibinfo{publisher}{ACM}, \bibinfo{address}{Santa Cruz CA USA}.
\newblock
\showISBNx{9798400710421}
\urldef\tempurl%
\url{https://doi.org/10.1145/3686169.3686179}
\showDOI{\tempurl}


\bibitem[Kreminski(2024b)]%
        {kreminski2024dearth}
\bibfield{author}{\bibinfo{person}{Max Kreminski}.} \bibinfo{year}{2024}\natexlab{b}.
\newblock \showarticletitle{The dearth of the author in AI-supported writing}. In \bibinfo{booktitle}{\emph{Proceedings of the Third Workshop on Intelligent and Interactive Writing Assistants}}. \bibinfo{pages}{48--50}.
\newblock


\bibitem[Kreminski and Chung(2024)]%
        {kreminski2024intent}
\bibfield{author}{\bibinfo{person}{Max Kreminski} {and} \bibinfo{person}{John Joon~Young Chung}.} \bibinfo{year}{2024}\natexlab{}.
\newblock \showarticletitle{Intent Elicitation in Mixed-Initiative Co-Creativity}. In \bibinfo{booktitle}{\emph{IUI Workshops}}.
\newblock


\bibitem[Kreminski and Martens(2022)]%
        {UnmetNeeds}
\bibfield{author}{\bibinfo{person}{Max Kreminski} {and} \bibinfo{person}{Chris Martens}.} \bibinfo{year}{2022}\natexlab{}.
\newblock \showarticletitle{Unmet creativity support needs in computationally supported creative writing}. In \bibinfo{booktitle}{\emph{Proceedings of the First Workshop on Intelligent and Interactive Writing Assistants (In2Writing 2022)}}. \bibinfo{pages}{74--82}.
\newblock


\bibitem[Kreminski and Mateas(2021a)]%
        {kreminski_reflective_nodate}
\bibfield{author}{\bibinfo{person}{Max Kreminski} {and} \bibinfo{person}{Michael Mateas}.} \bibinfo{year}{2021}\natexlab{a}.
\newblock \showarticletitle{Reflective {Creators}}. In \bibinfo{booktitle}{\emph{International Conference on Computational Creativity}}.
\newblock


\bibitem[Kreminski and Mateas(2021b)]%
        {mitchell_toward_2021}
\bibfield{author}{\bibinfo{person}{Max Kreminski} {and} \bibinfo{person}{Michael Mateas}.} \bibinfo{year}{2021}\natexlab{b}.
\newblock \showarticletitle{Toward {Narrative} {Instruments}}.
\newblock In \bibinfo{booktitle}{\emph{Interactive {Storytelling}}}, \bibfield{editor}{\bibinfo{person}{Alex Mitchell} {and} \bibinfo{person}{Mirjam Vosmeer}} (Eds.). Vol.~\bibinfo{volume}{13138}. \bibinfo{publisher}{Springer International Publishing}, \bibinfo{address}{Cham}, \bibinfo{pages}{499--508}.
\newblock
\showISBNx{978-3-030-92299-3 978-3-030-92300-6}
\urldef\tempurl%
\url{https://doi.org/10.1007/978-3-030-92300-6_50}
\showDOI{\tempurl}
\newblock
\shownote{Series Title: Lecture Notes in Computer Science}.


\bibitem[Lawton et~al\mbox{.}(2023)]%
        {Reframer}
\bibfield{author}{\bibinfo{person}{Tomas Lawton}, \bibinfo{person}{Kazjon Grace}, {and} \bibinfo{person}{Francisco~J Ibarrola}.} \bibinfo{year}{2023}\natexlab{}.
\newblock \showarticletitle{When is a tool a tool? User perceptions of system agency in human--{AI} co-creative drawing}. In \bibinfo{booktitle}{\emph{Proceedings of the 2023 ACM Designing Interactive Systems Conference}}. \bibinfo{pages}{1978--1996}.
\newblock


\bibitem[Lee et~al\mbox{.}(2024)]%
        {lee_design_2024}
\bibfield{author}{\bibinfo{person}{Mina Lee}, \bibinfo{person}{Katy~Ilonka Gero}, \bibinfo{person}{John Joon~Young Chung}, \bibinfo{person}{Simon~Buckingham Shum}, \bibinfo{person}{Vipul Raheja}, \bibinfo{person}{Hua Shen}, \bibinfo{person}{Subhashini Venugopalan}, \bibinfo{person}{Thiemo Wambsganss}, \bibinfo{person}{David Zhou}, \bibinfo{person}{Emad~A. Alghamdi}, \bibinfo{person}{Tal August}, \bibinfo{person}{Avinash Bhat}, \bibinfo{person}{Madiha~Zahrah Choksi}, \bibinfo{person}{Senjuti Dutta}, \bibinfo{person}{Jin L.~C. Guo}, \bibinfo{person}{Md~Naimul Hoque}, \bibinfo{person}{Yewon Kim}, \bibinfo{person}{Simon Knight}, \bibinfo{person}{Seyed~Parsa Neshaei}, \bibinfo{person}{Agnia Sergeyuk}, \bibinfo{person}{Antonette Shibani}, \bibinfo{person}{Disha Shrivastava}, \bibinfo{person}{Lila Shroff}, \bibinfo{person}{Jessi Stark}, \bibinfo{person}{Sarah Sterman}, \bibinfo{person}{Sitong Wang}, \bibinfo{person}{Antoine Bosselut}, \bibinfo{person}{Daniel Buschek}, \bibinfo{person}{Joseph~Chee Chang},
  \bibinfo{person}{Sherol Chen}, \bibinfo{person}{Max Kreminski}, \bibinfo{person}{Joonsuk Park}, \bibinfo{person}{Roy Pea}, \bibinfo{person}{Eugenia~H. Rho}, \bibinfo{person}{Shannon~Zejiang Shen}, {and} \bibinfo{person}{Pao Siangliulue}.} \bibinfo{year}{2024}\natexlab{}.
\newblock \showarticletitle{A {Design} {Space} for {Intelligent} and {Interactive} {Writing} {Assistants}}. In \bibinfo{booktitle}{\emph{Proceedings of the {CHI} {Conference} on {Human} {Factors} in {Computing} {Systems}}}.
\newblock
\urldef\tempurl%
\url{https://doi.org/10.1145/3613904.3642697}
\showDOI{\tempurl}


\bibitem[Li et~al\mbox{.}(2023)]%
        {li_beyond_2023}
\bibfield{author}{\bibinfo{person}{Jingyi Li}, \bibinfo{person}{Eric Rawn}, \bibinfo{person}{Jacob Ritchie}, \bibinfo{person}{Jasper Tran~O'Leary}, {and} \bibinfo{person}{Sean Follmer}.} \bibinfo{year}{2023}\natexlab{}.
\newblock \showarticletitle{Beyond the {Artifact}: {Power} as a {Lens} for {Creativity} {Support} {Tools}}. In \bibinfo{booktitle}{\emph{Proceedings of the 36th {Annual} {ACM} {Symposium} on {User} {Interface} {Software} and {Technology}}}. \bibinfo{publisher}{ACM}, \bibinfo{address}{San Francisco CA USA}, \bibinfo{pages}{1--15}.
\newblock
\showISBNx{9798400701320}
\urldef\tempurl%
\url{https://doi.org/10.1145/3586183.3606831}
\showDOI{\tempurl}


\bibitem[Long et~al\mbox{.}(2024)]%
        {long_not_2024}
\bibfield{author}{\bibinfo{person}{Tao Long}, \bibinfo{person}{Katy~Ilonka Gero}, {and} \bibinfo{person}{Lydia~B Chilton}.} \bibinfo{year}{2024}\natexlab{}.
\newblock \showarticletitle{Not {Just} {Novelty}: {A} {Longitudinal} {Study} on {Utility} and {Customization} of an {AI} {Workflow}}. In \bibinfo{booktitle}{\emph{Designing {Interactive} {Systems} {Conference}}}. \bibinfo{publisher}{ACM}, \bibinfo{address}{IT University of Copenhagen Denmark}, \bibinfo{pages}{782--803}.
\newblock
\showISBNx{9798400705830}
\urldef\tempurl%
\url{https://doi.org/10.1145/3643834.3661587}
\showDOI{\tempurl}


\bibitem[Mikolov et~al\mbox{.}(2013)]%
        {word2vec}
\bibfield{author}{\bibinfo{person}{Tomas Mikolov}, \bibinfo{person}{Kai Chen}, \bibinfo{person}{Greg Corrado}, {and} \bibinfo{person}{Jeffrey Dean}.} \bibinfo{year}{2013}\natexlab{}.
\newblock \showarticletitle{Efficient estimation of word representations in vector space}.
\newblock \bibinfo{journal}{\emph{arXiv preprint arXiv:1301.3781}} (\bibinfo{year}{2013}).
\newblock


\bibitem[Moretti(2005)]%
        {moretti_graphs_2005}
\bibfield{author}{\bibinfo{person}{Franco Moretti}.} \bibinfo{year}{2005}\natexlab{}.
\newblock \bibinfo{booktitle}{\emph{Graphs, maps, trees: abstract models for a literary history}}.
\newblock \bibinfo{publisher}{Verso}.
\newblock


\bibitem[Navigli et~al\mbox{.}(2021)]%
        {BabelNet}
\bibfield{author}{\bibinfo{person}{Roberto Navigli}, \bibinfo{person}{Michele Bevilacqua}, \bibinfo{person}{Simone Conia}, \bibinfo{person}{Dario Montagnini}, \bibinfo{person}{Francesco Cecconi}, {et~al\mbox{.}}} \bibinfo{year}{2021}\natexlab{}.
\newblock \showarticletitle{Ten years of {BabelNet}: A survey}. In \bibinfo{booktitle}{\emph{IJCAI}}. International Joint Conferences on Artificial Intelligence Organization, \bibinfo{pages}{4559--4567}.
\newblock


\bibitem[Padmakumar and He(2023)]%
        {PadmakumarHe}
\bibfield{author}{\bibinfo{person}{Vishakh Padmakumar} {and} \bibinfo{person}{He He}.} \bibinfo{year}{2023}\natexlab{}.
\newblock \showarticletitle{Does Writing with Language Models Reduce Content Diversity?}
\newblock \bibinfo{journal}{\emph{arXiv preprint arXiv:2309.05196}} (\bibinfo{year}{2023}).
\newblock


\bibitem[Parrish(2015)]%
        {parish_exploring_nodate}
\bibfield{author}{\bibinfo{person}{Allison Parrish}.} \bibinfo{year}{2015}\natexlab{}.
\newblock \bibinfo{title}{Exploring ({Semantic}) {Space} {With} ({Literal}) {Robots}}.
\newblock
\newblock
\urldef\tempurl%
\url{http://opentranscripts.org/transcript/semantic-space-literal-robots/}
\showURL{%
\tempurl}


\bibitem[Ramsay(2011)]%
        {ramsay_reading_2011}
\bibfield{author}{\bibinfo{person}{Stephen Ramsay}.} \bibinfo{year}{2011}\natexlab{}.
\newblock \bibinfo{booktitle}{\emph{Reading {Machines}: {Toward} and {Algorithmic} {Criticism}}}.
\newblock \bibinfo{publisher}{University of Illinois Press}.
\newblock


\bibitem[Robinson(2022)]%
        {robinson_speculative_2022}
\bibfield{author}{\bibinfo{person}{Bradley Robinson}.} \bibinfo{year}{2022}\natexlab{}.
\newblock \showarticletitle{Speculative {Propositions} for {Digital} {Writing} {Under} the {New} {Autonomous} {Model} of {Literacy}}.
\newblock \bibinfo{journal}{\emph{Postdigital Science and Education}} \bibinfo{volume}{5}, \bibinfo{number}{1} (\bibinfo{date}{Nov.} \bibinfo{year}{2022}), \bibinfo{pages}{117--135}.
\newblock
\showISSN{2524-485X, 2524-4868}
\urldef\tempurl%
\url{https://doi.org/10.1007/s42438-022-00358-5}
\showDOI{\tempurl}


\bibitem[Roemmele(2021)]%
        {roemmele2021inspiration}
\bibfield{author}{\bibinfo{person}{Melissa Roemmele}.} \bibinfo{year}{2021}\natexlab{}.
\newblock \showarticletitle{Inspiration through observation: Demonstrating the influence of automatically generated text on creative writing}.
\newblock \bibinfo{journal}{\emph{arXiv preprint arXiv:2107.04007}} (\bibinfo{year}{2021}).
\newblock


\bibitem[Roush et~al\mbox{.}(2022)]%
        {roush_most_2022}
\bibfield{author}{\bibinfo{person}{Allen Roush}, \bibinfo{person}{Sanjay Basu}, \bibinfo{person}{Akshay Moorthy}, {and} \bibinfo{person}{Dmitry Dubovoy}.} \bibinfo{year}{2022}\natexlab{}.
\newblock \showarticletitle{Most Language Models can be Poets too: An {AI} Writing Assistant and Constrained Text Generation Studio}. In \bibinfo{booktitle}{\emph{Proceedings of the Second Workshop on When Creative AI Meets Conversational AI}}, \bibfield{editor}{\bibinfo{person}{Xianchao Wu}, \bibinfo{person}{Peiying Ruan}, \bibinfo{person}{Sheng Li}, {and} \bibinfo{person}{Yi~Dong}} (Eds.). \bibinfo{publisher}{Association for Computational Linguistics}, \bibinfo{address}{Gyeongju, Republic of Korea}, \bibinfo{pages}{9--15}.
\newblock
\urldef\tempurl%
\url{https://aclanthology.org/2022.cai-1.2/}
\showURL{%
\tempurl}


\bibitem[Singh et~al\mbox{.}(2022)]%
        {singh_where_2022}
\bibfield{author}{\bibinfo{person}{Nikhil Singh}, \bibinfo{person}{Guillermo Bernal}, \bibinfo{person}{Daria Savchenko}, {and} \bibinfo{person}{Elena~L. Glassman}.} \bibinfo{year}{2022}\natexlab{}.
\newblock \showarticletitle{Where to hide a stolen elephant: {Leaps} in creative writing with multimodal machine intelligence}.
\newblock \bibinfo{journal}{\emph{ACM Transactions on Computer-Human Interaction}} (\bibinfo{year}{2022}).
\newblock
\newblock
\shownote{Publisher: ACM New York, NY}.


\bibitem[Stark et~al\mbox{.}(2023)]%
        {WithoutWriting2}
\bibfield{author}{\bibinfo{person}{Jessi Stark}, \bibinfo{person}{Anthony Tang}, \bibinfo{person}{Young-Ho Kim}, \bibinfo{person}{Joonsuk Park}, {and} \bibinfo{person}{Daniel Wigdor}.} \bibinfo{year}{2023}\natexlab{}.
\newblock \showarticletitle{Can {AI} Support Fiction Writers Without Writing For Them?\killpunct}. In \bibinfo{booktitle}{\emph{Proceedings of the Second Workshop on Intelligent and Interactive Writing Assistants}}.
\newblock


\bibitem[Sullivan(2013)]%
        {sullivan_work_2013}
\bibfield{author}{\bibinfo{person}{Hannah Sullivan}.} \bibinfo{year}{2013}\natexlab{}.
\newblock \bibinfo{booktitle}{\emph{The {Work} of {Revision}}}.
\newblock \bibinfo{publisher}{Harvard University Press}.
\newblock
\showISBNx{978-0-674-07312-8}
\newblock
\shownote{Google-Books-ID: gC\_WEAAAQBAJ}.


\bibitem[Sun et~al\mbox{.}(2024)]%
        {Persona-L}
\bibfield{author}{\bibinfo{person}{Lipeipei Sun}, \bibinfo{person}{Tianzi Qin}, \bibinfo{person}{Anran Hu}, \bibinfo{person}{Jiale Zhang}, \bibinfo{person}{Shuojia Lin}, \bibinfo{person}{Jianyan Chen}, \bibinfo{person}{Mona Ali}, {and} \bibinfo{person}{Mirjana Prpa}.} \bibinfo{year}{2024}\natexlab{}.
\newblock \showarticletitle{Persona-L has Entered the Chat: Leveraging LLM and Ability-based Framework for Personas of People with Complex Needs}.
\newblock \bibinfo{journal}{\emph{arXiv preprint arXiv:2409.15604}} (\bibinfo{year}{2024}).
\newblock


\bibitem[Tanaka(2006)]%
        {tanaka2006interaction}
\bibfield{author}{\bibinfo{person}{Atau Tanaka}.} \bibinfo{year}{2006}\natexlab{}.
\newblock \showarticletitle{Interaction, experience and the future of music}.
\newblock In \bibinfo{booktitle}{\emph{Consuming music together: Social and collaborative aspects of music consumption technologies}}. \bibinfo{publisher}{Springer}, \bibinfo{pages}{267--288}.
\newblock


\bibitem[Thomas(1988)]%
        {thomas_oulipo_1988}
\bibfield{author}{\bibinfo{person}{Jean-Jacques Thomas}.} \bibinfo{year}{1988}\natexlab{}.
\newblock \bibinfo{title}{Oulipo: {A} {Primer} of {Potential} {Literature}}.
\newblock
\newblock


\bibitem[Wardrip-Fruin(2007)]%
        {wardrip_zot}
\bibfield{author}{\bibinfo{person}{Noah Wardrip-Fruin}.} \bibinfo{year}{2007}\natexlab{}.
\newblock \showarticletitle{Playable media and textual instruments}.
\newblock \bibinfo{journal}{\emph{The aesthetics of net literature: writing, reading and playing in programmable media}} (\bibinfo{year}{2007}), \bibinfo{pages}{211--80}.
\newblock


\bibitem[Yuan et~al\mbox{.}(2022)]%
        {wordcraft}
\bibfield{author}{\bibinfo{person}{Ann Yuan}, \bibinfo{person}{Andy Coenen}, \bibinfo{person}{Emily Reif}, {and} \bibinfo{person}{Daphne Ippolito}.} \bibinfo{year}{2022}\natexlab{}.
\newblock \showarticletitle{Wordcraft: story writing with large language models}. In \bibinfo{booktitle}{\emph{Proceedings of the 27th International Conference on Intelligent User Interfaces}}. \bibinfo{pages}{841--852}.
\newblock


\bibitem[Zhou and Sterman(2024)]%
        {zhou2024ai}
\bibfield{author}{\bibinfo{person}{David Zhou} {and} \bibinfo{person}{Sarah Sterman}.} \bibinfo{year}{2024}\natexlab{}.
\newblock \showarticletitle{{Ai.llude}: Investigating Rewriting {AI}-Generated Text to Support Creative Expression}. In \bibinfo{booktitle}{\emph{Proceedings of the 16th Conference on Creativity \& Cognition}}. \bibinfo{pages}{241--254}.
\newblock


\bibitem[Zhou et~al\mbox{.}(2022)]%
        {zhou_maintenance_2022}
\bibfield{author}{\bibinfo{person}{Hongwei Zhou}, \bibinfo{person}{Kyle Gonzalez}, \bibinfo{person}{Nathan Altice}, {and} \bibinfo{person}{Angus~G Forbes}.} \bibinfo{year}{2022}\natexlab{}.
\newblock \showarticletitle{On the {Maintenance} of {Meaning}: {A} {Deleuzian} {View} on {Proceduralism}}. In \bibinfo{booktitle}{\emph{Proceedings of DiGRA 2022}}.
\newblock


\end{thebibliography}
\bibliographystyle{ACM-Reference-Format}

\onecolumn  
\appendix
\section{(Self Reported) Participant Demographics} \label{demographics}
\begin{table}[htbp]  
\small
\rowcolors{2}{gray!10}{white} 
\begin{tabular}{p{0.12\textwidth}p{0.38\textwidth}p{0.42\textwidth}}
\toprule
Participant ID & Demographics & Writing Practice \\
\midrule
P1 & 33 cis-male he/him writer from New Zealand & publishing for 6 years, multi-genre - poetry, CNF, fiction \\
P2 & 22, he/him, Black/African, grew up in Nigeria & poet, interested in memory and history, publishing for four years. \\
P3 & 33, white cis male, he/him, American & poet (both experimental and traditional), musician, MFA in poetry, publishing for 13 years \\
P4 & 28, male & 6 years of publishing experience. \\
P5 & 40, she/her, Chinese-American, multilingual & poet \& creative writing instructor, publishing for 8 years \\
P6 & 30, they/them, androgynous & poet turned prose writer, Ph.D. student. \\
P7 & mid-30s, Chinese American woman, they/them & publishing for 6 years \\
P8 & 30, he/him, white, American & neosymbolist (idk) poet and translator, publishing for 5 years. \\
P9 & 53, she/her, cis female, white & I've been writing poetry for over thirty years, publishing over twenty. I write toward poems more days than not, even if just journaling or taking notes. I do go through dry spells, which I use to purposefully gather experiences, including wider reading. \\
P10 & European French/English ethnicity, cis female, grandmother, activist, immigrant to Canada & poet, editor, essayist, mentor, artistic collaborator. \\
\bottomrule
\end{tabular}
\caption{Self Reported Participant Demographics and Writing Practices}
\label{tab:participants}
\end{table}

\end{document}